\documentclass[aps,preprint,groupedaddress,floats]{revtex4}
\usepackage{amsfonts}
\usepackage[final]{graphicx}
\usepackage{feynmf}

\begin{document}
\bibliographystyle{revtex}


\newcommand{\alps}{\ensuremath{\alpha_s}}
\newcommand{\qbar}{\bar{q}}
\newcommand{\beq}{\begin{equation}}
\newcommand{\eeq}{\end{equation}}
\newcommand{\mq}{m_Q}
\newcommand{\mn}{m_N}
\newcommand{\bb}{\langle}
\newcommand{\kb}{\rangle}
\newcommand{\st}{\ensuremath{\sqrt{\sigma}}}
\newcommand{\rvec}{\mathbf{r}}
\newcommand{\bvec}[1]{\ensuremath{\mathbf{#1}}}
\newcommand{\bra}[1]{\ensuremath{\bb#1|}}
\newcommand{\ket}[1]{\ensuremath{|#1\kb}}
\newcommand{\gft}{\ensuremath{\gamma_{FT}}}
\newcommand{\bfsig}{\mbox{\boldmath{$\sigma$}}}
\newcommand{\bfnab}{\mbox{\boldmath{$\nabla$}}}
\newcommand{\bftau}{\mbox{\boldmath{$\tau$}}}
\newcommand{\spup}{\uparrow}
\newcommand{\spd}{\downarrow}
\newcommand{\hbarom}{\frac{\hbar^2}{m_Q}}
\newcommand{\vnn}{\ensuremath{\hat{v}_{NN}}}
\newcommand{\argonne}{\ensuremath{v_{18}}}

\title[QMC calculations of six quarks]
{Quantum Monte Carlo calculations of six-quark states}

\author{Mark W. Paris}
\email[]{paris@uiuc.edu}
\author{Vijay R. Pandharipande}
\email[]{vrp@uiuc.edu}
\affiliation{Department of Physics,
University of Illinois at Urbana-Champaign,
1110 West Green Street, Urbana, Illinois 61801}

\date{\today}

\begin{abstract}
The variational Monte Carlo method is used to find the ground
state of six quarks confined to a cavity of diameter $R_c$,
interacting via an assumed non-relativistic constituent
quark model (CQM) Hamiltonian. We use a flux-tube
model augmented with one-gluon and one-pion exchange interactions,
which has been successful in describing single hadron spectra.
The variational wave function is written as a product of
three-quark nucleon states with correlations between quarks
in different nucleons. We study the role of quark exchange effects
by allowing flux-tube configuration mixing. An accurate six-body
variational wave function is obtained. It has only $\sim 13\%$
rms fluctuation in the total energy and yields a
standard deviation of $\lesssim .1\%$; small enough to be useful
in discerning nuclear interaction effects from the large rest mass of
the two nucleons. Results are presented for three values of the
cavity diameter, $R_c=2$, 4, and 6 fm. They indicate that the flux-tube
model Hamiltonian with gluon and pion exchange requires
revisions in order to obtain agreement with the energies estimated
from realistic two-nucleon interactions.
We calculate the two-quark probability distribution functions and
show how they may be used to study and adjust the model Hamiltonian.
\end{abstract}
\maketitle

\section{Introduction}
\label{intro}
The constituent quark model (CQM) has been useful to describe the
spectroscopy \cite{isgur:3q, robson:3q, ckp2} and the decay
\cite{isgur:decay, stancu, kumano1} of baryons and mesons.
It retains only the constituent quark (CQ) degrees-of-freedom; the
effects of all other degrees-of-freedom are subsumed into potentials
dependent upon CQ positions, spins, flavors and colors. It assumes that a
Hamiltonian describing interacting quarks can provide a useful
description of the low-energy properties of hadrons.

It is natural to attempt to extend the CQM to describe the
two-baryon states made up of six CQ. The only such bound state
known is the deuteron; all other known two-baryon states are unbound.
The scattering data provides information on the interaction between
baryons which is used to construct realistic models of the
two-baryon potential \cite{nijmegen, wiringa:av18, rijken}.

Many authors \cite{isgur:6q, oka84, robson:6q} have attempted to
calculate properties of two-baryon states, including the interaction
between nucleons, using the CQM. In these studies some approximation
scheme is used to avoid the calculation of six-body eigenstates of the
assumed CQM Hamiltonian. In the past few years there have been
significant advances in the variational (VMC) and Green's function
Monte Carlo (GFMC) methods to calculate the eigenstates of up to
eight interacting nucleons \cite{pudliner1, wiringa00}. In these
quantum Monte Carlo (QMC) methods a good approximate solution is
first obtained by the VMC method from which the exact eigenstate is
projected with the GFMC.

In the present work we attempt to find the ground state of two-nucleons
confined to a spherical cavity of diameter $R_c$ in the center-of-mass
frame using the CQM. In the interior of the cavity the six CQ wave
function is determined with VMC. Near the edge of the cavity we assume
that it factorizes into a two-nucleon wave function required to be zero
at the internucleon distance $R_c$.

This problem may be easily solved assuming that the ground state can
be described as that of two interacting nucleons. The deuteron has
total spin-isopin, $S,T=1,0$, and in local, realistic models such as
Argonne \argonne\ \cite{wiringa:av18}, the nuclear interaction
in the deuteron is given by a sum of
central, tensor, and spin-orbit potentials: $v_d^c(r)$,
$v_d^t(r)$ and $v_d^{\ell s}(r)$. The two-nucleon Schr\"{o}dinger equation:
\begin{eqnarray}
\label{eqn:urel}
-\frac{\hbar^2}{m_N} \frac{d^2}{dr^2}u(r) + v_d^c(r) u(r)
+\sqrt{8}v_d^t(r)w(r)&=&E(R_c)u(r) \\
\label{eqn:wrel}
-\frac{\hbar^2}{m_N} \left( \frac{d^2}{dr^2}-\frac{6}{r^2} \right) w(r)
+ \left( v_d^c(r)-2 v_d^t(r)-3 v_d^{\ell s}(r) \right) w(r) \nonumber \\
+ \sqrt{8} v_d^t(r) u(r)&=&E(R_c) w(r)
\end{eqnarray}
where $m_N$ is the average mass of the neutron and proton,
and $u(r)$ and $w(r)$, the ${^3S}_1$ and ${^3D}_1$ wave
functions, can be easily solved with the boundary conditions
$u(R_c)=w(R_c)=0$. The results obtained with the \argonne\ model are
shown in Table \ref{tab:av18} for $R_c =$ 2, 4, and 6 fm. They
presumably have small dependence on the model $NN$ interaction because
all models have mostly the one-pion exchange potential (OPEP),
$v^\pi_{NN}$
at $r>2$ fm in $T=0$ states. Note that $E(R_c\to\infty)$ is
just the deuteron binding energy of $-2.24$ MeV, for the isoscalar
part of Argonne \argonne.

The $E(R_c)=\bb T_N \kb + \bb v_{NN} \kb$, has a large cancellation
between the nucleon kinetic energy $\bb T_N \kb$ and the negative
two-nucleon interaction energy $\bb v_{NN} \kb$, typical of nuclear
systems \cite{pudliner1}. In Table \ref{tab:av18} we also list the
energy,
\beq
\label{eqn:freeNN}
E_{NI}(R_c) = \frac{\pi^2\hbar^2}{m_N R_c^2},
\eeq
of two non-interacting nucleons in a cavity of diameter $R_c$ and the
total effect $\delta E_{emp}(R_c) = E(R_c) - E_{NI}(R_c)$ of strong
interactions, per nucleon.  Our long-range aim is to calculate
$E(R_c)$, and thus $\delta E(R_c)$, with computational errors
of $<1$ MeV using CQM, and compare with the
$\delta E_{emp}(R_c)$, calculated from Argonne \argonne\ 
potential, which reproduces the known $NN$ data. These
calculations can then be used to refine the CQM Hamiltonian and
study the behavior of the wave function in the region where the six
CQ are close together, i.e.\ the region in which the quark distributions
of the two-nucleons overlap. Eventually such calculations may be useful
to study $\Lambda -N$ and $\Sigma -N$ interactions for which the
scattering data is limited.

The first main concern in this approach is that the total energy,
$E_V$, in the CQM
includes the rest mass of the nucleons, and is therefore of order
$2m_N \sim$ 2000 MeV. In order to calculate this energy with an
accuracy of $\sim 1$ MeV the variance of the local energy,
\beq
\label{eqn:localen}
E_V(\bvec{R})=\frac{\Psi_V^{\dagger}(\bvec{R})\hat{H} \Psi_V(\bvec{R}) }
{ \Psi_V^{\dagger}(\bvec{R}) \Psi_V(\bvec{R})},
\eeq
where $\bvec{R}=\{\bvec{r}_1, \bvec{r}_2, \dots ,\bvec{r}_6 \}$ is
the six CQ configuration and $\Psi_V(\bvec{R})$ is the
variational wave function, must be sufficiently small so that the
Monte Carlo statistical error is of $\lesssim 1$ MeV. A major focus of
this work is on developing six CQ wave functions with small variance
of the local energy. The exact eigenstate satisfies
$\hat{H} \Psi_0(\bvec{R}) = E_0 \Psi_0(\bvec{R})$ at all $\bvec{R}$,
and hence has zero variance. The Monte Carlo statistical errors in
GFMC calculations are also determined by the variance of
$E_V(\bvec{R})$, thus it is necessary to obtain variational wave
functions with small variance before proceeding to GFMC.

The second main concern is the choice of the CQM Hamiltonian. Our
$\hat{H}$ is basically a generalization of the Y-junction flux-tube (FT)
model \cite{ckp2,ckp1}. The model Hamiltonian in Ref.\cite{ckp2}
(herein referred to as CKP) contains
relativistic CQ kinetic energies, Y-junction FT confinement
potential and one-gluon exchange Coulomb, $\bfsig_i\cdot\bfsig_j$,
tensor and spin-orbit interactions. It is not suitable to study 
six-quark systems for at least two reasons.

Any model Hamiltonian used for six light quarks must be able to
describe two free nucleons. In order to do that
it must give $E(\bvec{p})=m_N + p^2/(2m_N)$ for the energy of a
nucleon with momentum $p\ll m_N$. This requirement is easily acheived
in non-relativistic Hamiltonians by choosing the mass $m_Q$ of the
CQ as $m_N/3$. In the Hamiltonians containing relativistic kinetic
energies it is necessary to include boost corrections
\cite{carlson:rel, forest:rel1} to all interactions
to satisfy this requirement. These are absent in the CKP Hamiltonian.
In order to avoid them we begin with the non-relativistic model.
However, the CQ momenta are often greater
than $m_N/3$, and it will be necessary to work with the relativistic
Hamiltonian to obtain more reliable results. The present calculation
is just the first step.

Second, it is well known that nucleons a few fermi apart interact
via the OPEP. It has been shown that the tensor force in the OPEP
largely determines the toroidal structure of the deuteron 
\cite{donuts}. The model Hamiltonian of CKP does
not contain pion-exchange interactions. Within that approach, the
emission and absorption of pions by nucleons may be
attributed to breaking of flux-tubes \cite{kumano2}, an
effect absent in our present work. The simplest way to include such
processes in our CQM, which we adopt, is by coupling pion fields to
the CQ.

The advantages of replacing the one-gluon exchange (OGE) interaction
in CQM by OPE, as well as the problems associated with this have
been recently discussed by Glozman and Riska \cite{riska96, glozman99} and 
by Isgur \cite{isgur99}. In our model we include both OGE and OPE
interactions between the CQ. They respectively respresent the effect
of virtual gluons and color-singlet $q\qbar$ pairs missing in the CQM
Hamiltonian, and both should be included. A more realistic CQM
Hamiltonian may contain additional meson-exchange terms, as discussed
in Sec.\ref{sec:results}.

The details of the CQM Hamiltonian used in this work are given in
Sec.\ref{cqm}, and the variational wave function and VMC calculations
are described in Sec.\ref{vmc}. The results, indicating the
limitations of the present Hamiltonian, are presented and discussed
in the last section, Sec.\ref{sec:results}. Continuing research and
conclusions are given in the last Sec.\ref{sec:concl}.

\section{Flux-tube constituent quark model Hamiltonian}
\label{cqm}
\subsection{Single nucleon Hamiltonian}
\label{sec:H_N}
The Hamiltonian for the single nucleon is written as a sum
of kinetic energy and two and three-body potential operators,
\beq
\hat{H}_N = \sum_{i=1}^{3} \left( m_Q + {\hat{\bvec{p}}_i^2\over 2m_Q} \right)
+ \sum_{i<j=1}^{3} \hat{v}_{ij} + V_3 - V_0.
\label{eqn:H_N}
\eeq
The two-body term, $\hat{v}_{ij}$ is summed over pairs $(ij)$,
and $V_3$ is a three-body interaction. These potentials are due to
FT confinement, and perturbative OGE and OPE between constituent quarks.
The $V_0$ includes all constants in the long range confining 
interaction and is taken as a parameter to fit
the mass of the nucleon. In the case of the single nucleon in its 
rest frame, the constituent quark mass,
$m_Q$, is simply a parameter which may be arbitrarily chosen to
obtain agreement with experimental data.
However, in the present non-relativistic $6Q$ Hamiltonian
the constituent quark mass 
must be one-third of the nucleon mass for the Hamiltonian to 
reduce to that of two free nucleons when two three-quark clusters are
far apart.  For this reason we use $m_Q = m_N/3$ in this work.  

\begin{figure}[hb]
\begin{center}
\includegraphics[ width= 340 pt, keepaspectratio, clip ]{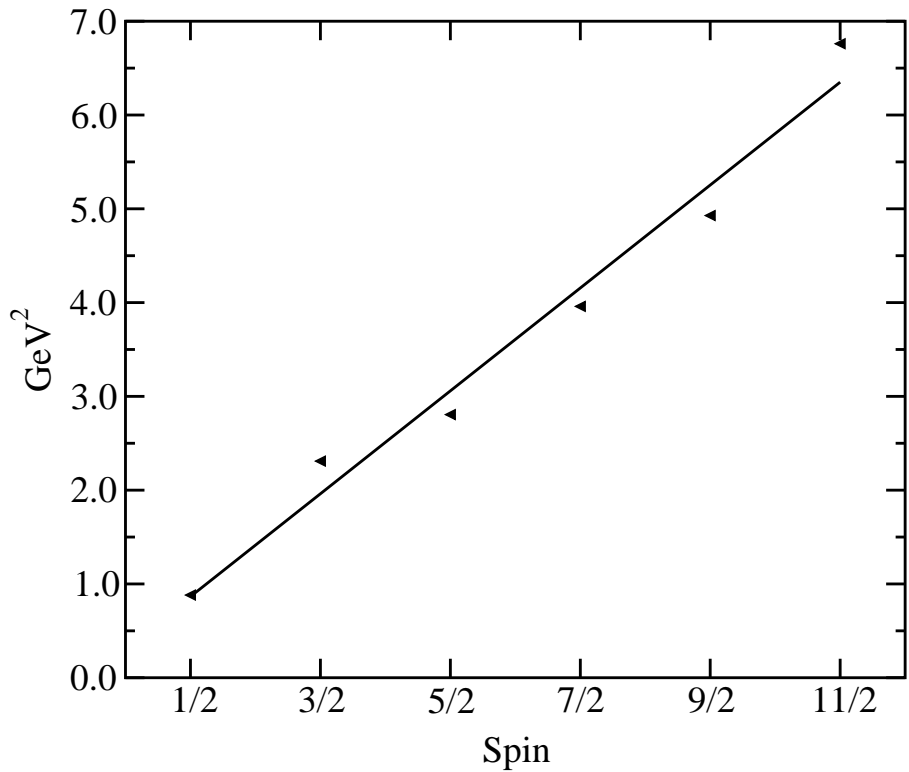}
\nopagebreak
\caption{\label{fig:regge}Nucleon mass squared versus $J$ in GeV$^2$.
The slope is reproduced with string tension $\st=0.88$ GeV.}
\end{center}
\end{figure}

We use the $SU(3)$ FT model of long-range confinement \cite{ckp1}.
The energy of the flux tubes connecting the three quarks is given by,
\beq
V^C(\rvec_1,\rvec_2,\rvec_3)=\st\sum_{i=1}^3 r_{iY},
\label{eqn:conf}
\eeq
where $r_{iY}$ is the length of the vector $\rvec_{iY}=\rvec_i-\rvec_Y$,
$\rvec_Y$ is the location of the ``Y-junction'' of the three flux-tubes
determined by the minimization of
$V^C$ with respect to $\rvec_Y$. The renormalized string tension, $\st$,
is fixed from the observed single-hadron Regge trajectories.
Following Johnson and Thorn\cite{johnson:regge} we assume,
\beq
\label{eqn:regge}
E_J^2 = 2\pi\st J \ ,
\eeq
where $E_J$ is the mass of the state with angular momentum $J$. 
The fit obtained for the nucleon trajectory is shown in 
Fig.(\ref{fig:regge}),
The slope gives $\st=0.88$ GeV/fm; a similar analysis
of $\Delta$ resonances produces essentially the same result. 
This value of $\st$, used here is smaller than the 1 GeV/fm 
used by CKP.

Considered
as quantum strings, the flux-tubes undergo zero-point oscillations which
give rise to a constant term, which is included in the $V_0$, 
and terms which depend inversely on the lengths $r_{iY}$ \cite{isgur:ftm}.
The latter are approximately merged into the color Coulomb potential.

We may break-up $V^C$ into two- and three-body terms,
\beq
V^C(\rvec_1,\rvec_2,\rvec_3)=
{1\over2} \st\sum_{i<j=1}^3 r_{ij} + V_{3} \ , 
\eeq
by defining the three-body potential:
\beq
V_{3}=\st\left(\sum_{i=1}^3 r_{iY}-{1\over2} \sum_{i<j=1}^3 r_{ij}\right).
\label{eqn:v3}
\eeq
This break-up is useful because the 
ratio of the three-body to the two-body term is $\le 0.154$
for all configurations, saturating when the quarks lie on the
vertices of an equilateral triangle. The $V_3$ term is zero
when the quarks are in a line, and approaches zero when one of the
quarks is far from the other two.

The FT confinement model is suggested by numerical studies of
lattice QCD which show the interaction between static quarks to be
linear in the quark separation at large distances and independent
of their spin and isospin \cite{stack}. The mechanism which gives
rise to a linear confining interaction was originally studied
by Wilson \cite{wilson} and others using path-integral methods.
Both lattice QCD and path-integral methods indicate that at
strong-coupling or, equivalently at length scales  
of order of the characteristic length scale of QCD,
believed to be $\sim 0.5$ fm \cite{marciano}, the
gluonic flux coalesces into tube or string-like configurations
due to an attractive self-interaction of the gluons. 

We also make the assumption that the flux-tubes move adiabatically,
the quarks remaining on their lowest-energy surface. The quarks
therefore determine the locations of the flux-tubes and neither the
flux nor the Y-junction are free dynamical variables. This seems to
be a rather good approximation since, as shown in Ref.\cite{page},
the lowest lying hybrid states, which take into account dynamics of
the flux-tubes, appear to be $\sim 1$ GeV above the nucleon. FT
topologies which have more than one Y-junction or loops of flux
correspond to higher energy surfaces \cite{isgur:ftm} and are ignored.

As the distance between sources gets very large
flux-tubes may break with concomitant quark-antiquark
pair creation. This process has been used to describe the
two-pion decays of mesons \cite{kumano1} as well as decays of
baryons \cite{kumano2}. The FT connected to a given quark
$i$ may break and the broken piece can reattach to the tube of a
different quark $j$. The effects of such interactions between
quarks $i$ and $j$ are to be included in the meson exchange terms
in CQM Hamiltonian. Of these we consider only the OPE in this work. 
The virtual $q\qbar$ creation also renormalizes
the string tension to its physical value \cite{isgur:ftm}. 

The total two-quark potential $\hat{v}_{ij}$
is given by the sum of confinement,
perturbative OGE and OPE potentials denoted by 
$\hat{v}^g_{ij}$ and $\hat{v}^\pi_{ij}$:
\beq
\label{eqn:v2}
\hat{v}_{ij} = {1 \over 2}\st \  r_{ij} + \hat{v}^g_{ij} + \hat{v}^\pi_{ij}.
\eeq
The OGE potential was first used in a CQM by DeR\'{u}jula, Georgi,
and Glashow \cite{DGG} who showed that the splitting within
single-hadron multiplets are qualitatively consistent with those 
expected from OGE interaction. The OGE potential is the
QCD analogue of the Fermi-Breit interaction in QED.

\begin{figure}[hb]
\begin{center}
\includegraphics[ width= 340 pt, keepaspectratio, clip ]{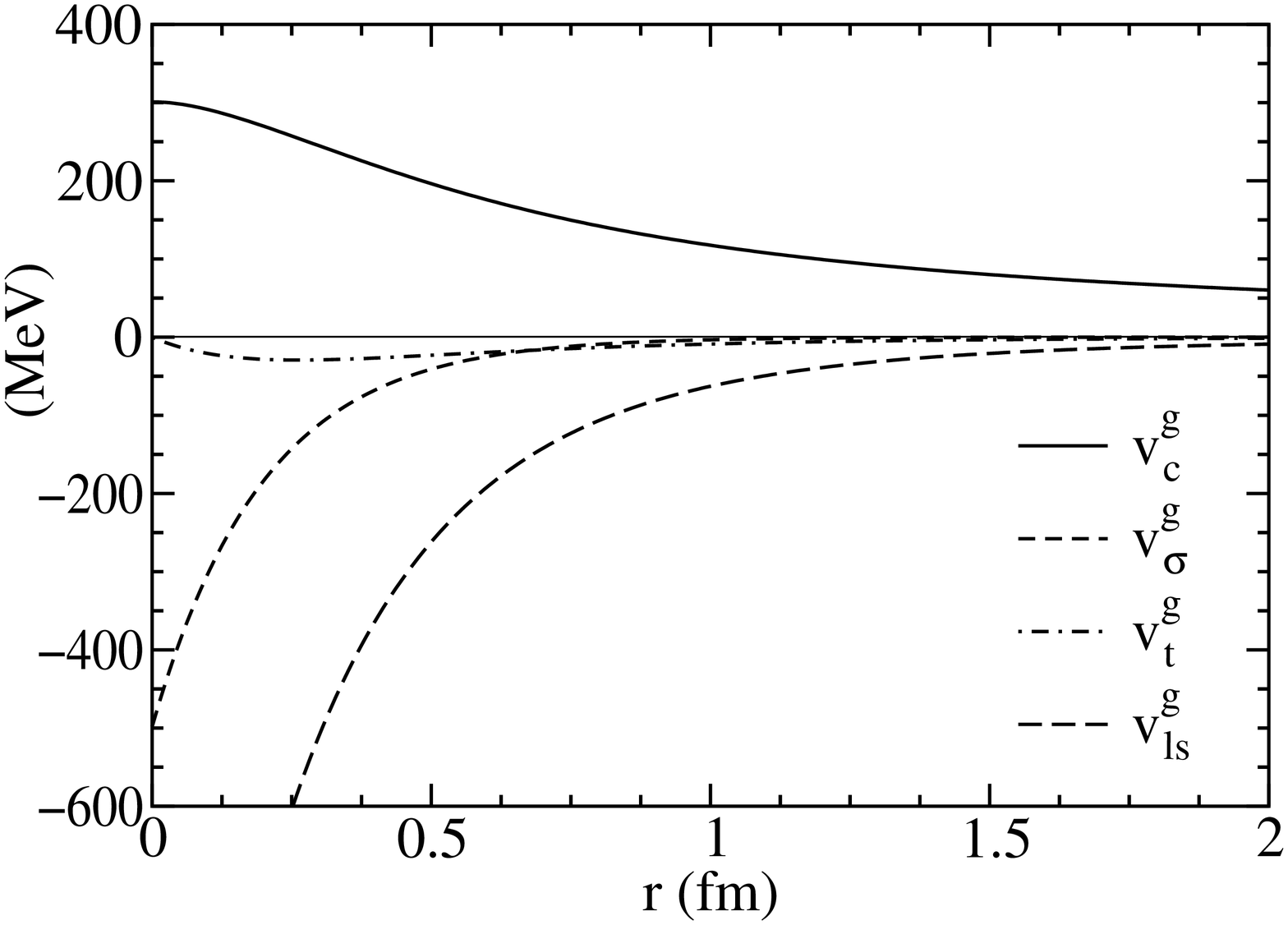}
\nopagebreak
\caption{\label{fig:vg} Regulated OGE potential in MeV.}
\end{center}
\end{figure}

For point particles the OGE spin interactions are singular in 
nonrelativistic theory. Taking into
account the finite  size of CQ, we may regulate the potentials
in order to obtain solutions of the Schr\"{o}dinger equation. We use
``monopole'' form factors at each vertex
\beq
\label{ff:mp}
F(\bvec{q}^2)={\Lambda^2\over\Lambda^2+\bvec{q}^2}
\eeq
where $\bvec{q}$ is the three-momentum transfer and $\Lambda$ is
the momentum space cutoff, which we take to be
5 fm$^{-1}$. The color-charge density of quarks is Yukawa-like
with this form factor. The $\hat{v}^g_{ij}$ is approximated by:
\beq
\hat{v}^g_{ij} = \sum_{p=c,\sigma,t,\ell s} v^g_p(r_{ij}) O^p_{ij}
                     \bvec{\mathsf{T}}_i\cdot\bvec{\mathsf{T}}_j
\label{eqn:oge}
\eeq
where the
$p=c$ term is the color-Coulomb interaction with $O^c_{ij}=\openone$;
$p=\sigma$ term denotes the color-magnetic contact interaction with 
$O^\sigma_{ij}=\bfsig_i\cdot\bfsig_j$; $t$ is for the tensor interaction,
$O^t_{ij}=S_{ij}=
3\bfsig_i\cdot\bvec{\hat{r}}_{ij}\bfsig_j\cdot\bvec{\hat{r}}_{ij}
-\bfsig_i\cdot\bfsig_j$; and $\ell s$ is the spin-orbit contribution,
$O^{\ell s}_{ij}=\bvec{L}_{ij}\cdot\bvec{S}_{ij}$. The
operator $\bvec{\mathsf{T}}_i\cdot\bvec{\mathsf{T}}_j$ is the quadratic
Casimir of $SU(3)$ acting on color indices of the quarks. The color wave 
function of three quark states is antisymmetric.  It is an eigenstate 
of $\bvec{\mathsf{T}}_i\cdot\bvec{\mathsf{T}}_j $ with eigenvalue $-2/3$. 
The specific forms of the monopole regulated potential functions:
\begin{eqnarray}
\label{eqn:vg1}
&v^g_c(r)& = {\alps\over r}
\left[1 - \left({\Lambda r\over 2}+1\right) e^{-\Lambda r}\right] \\
\label{eqn:vg2}
&v^g_\sigma(r)& = - {\alps \over 12 m_Q^2} \Lambda^3 e^{-\Lambda r} \\
\label{eqn:vg3}
&v^g_t(r)& = -{\alps \over 4 m_Q^2}
\left[{1\over r^3} - {\Lambda^3\over 3}\left(T_\Lambda(r)
+ {1\over 2}\left(\Lambda r+1\right) Y_\Lambda(r)\right)\right] \\
\label{eqn:vg4}
&v^g_{\ell s}(r)& = -{3\alps \over 2 m_Q^2}
\left[{1\over r^3} - {\Lambda^2\over r}
\left({1\over \Lambda r} + 1 + \Lambda r\right)Y_\Lambda(r)\right]
\end{eqnarray}
are plotted in Fig.(\ref{fig:vg}). The functions $Y_\Lambda$ and
$T_\Lambda$ are the Yukawa and tensor functions,
\begin{eqnarray}
\label{eqn:yukawa}
Y_\Lambda (r)&=&\frac{e^{-\Lambda r}}{\Lambda r}\\
\label{eqn:tensyuk}
T_\Lambda (r)&=&\left(1+\frac{3}{\Lambda r}+
\frac{3}{\Lambda^2 r^2}\right)Y_\Lambda (r)
\end{eqnarray}
The strength of the OGE potentials is given by the perturbative
strong coupling constant, \alps. We fix the value of $\alps=0.61$,
consistent with values in earlier works, to reproduce
the $N$--$\Delta$ splitting.

The OGE tensor and spin-orbit terms are not as well-established as
the color-Coulomb and the short-ranged spin-spin terms; the latter
gives a large contribution to the splitting of the pseudoscalar
mesons, $\pi-\rho$, and of the S-wave baryons, $N-\Delta$,
while the Coulomb term is seen in lattice QCD treatments of the
static $QQ$ potential \cite{stack}.

There is much discussion
in the literature concerning the apparent lack of the spin-orbit
term \cite{ckp2,isgur:qm,myrher} in nucleon 
spectra. Many workers simply discard the
longe-range, spin-dependent parts of OGE--the tensor and spin-orbit
forces--arguing that those terms may be significantly modified
by FT formation and are not well-supported by experimental data.
Isgur \cite{isgur:3q} has argued that the OGE spin-orbit term exists,
but it primarily cancels the neglected
spin-orbit interaction due to the Thomas 
precession in the confining FT potential. Oka and Yazaki \cite{oka84}
have shown that the tensor part of $\hat{v}^g_{ij}$ also gives a small
contribution to the baryon-baryon interactions. Neither the spin-orbit
nor tensor parts of $\hat{v}^g_{ij}$ seem to be important
in determining the $NN$ interaction in the state with deuteron
quantum numbers.

\begin{figure}[hb]
\begin{center}
\includegraphics[ width= 340 pt, keepaspectratio, clip ]{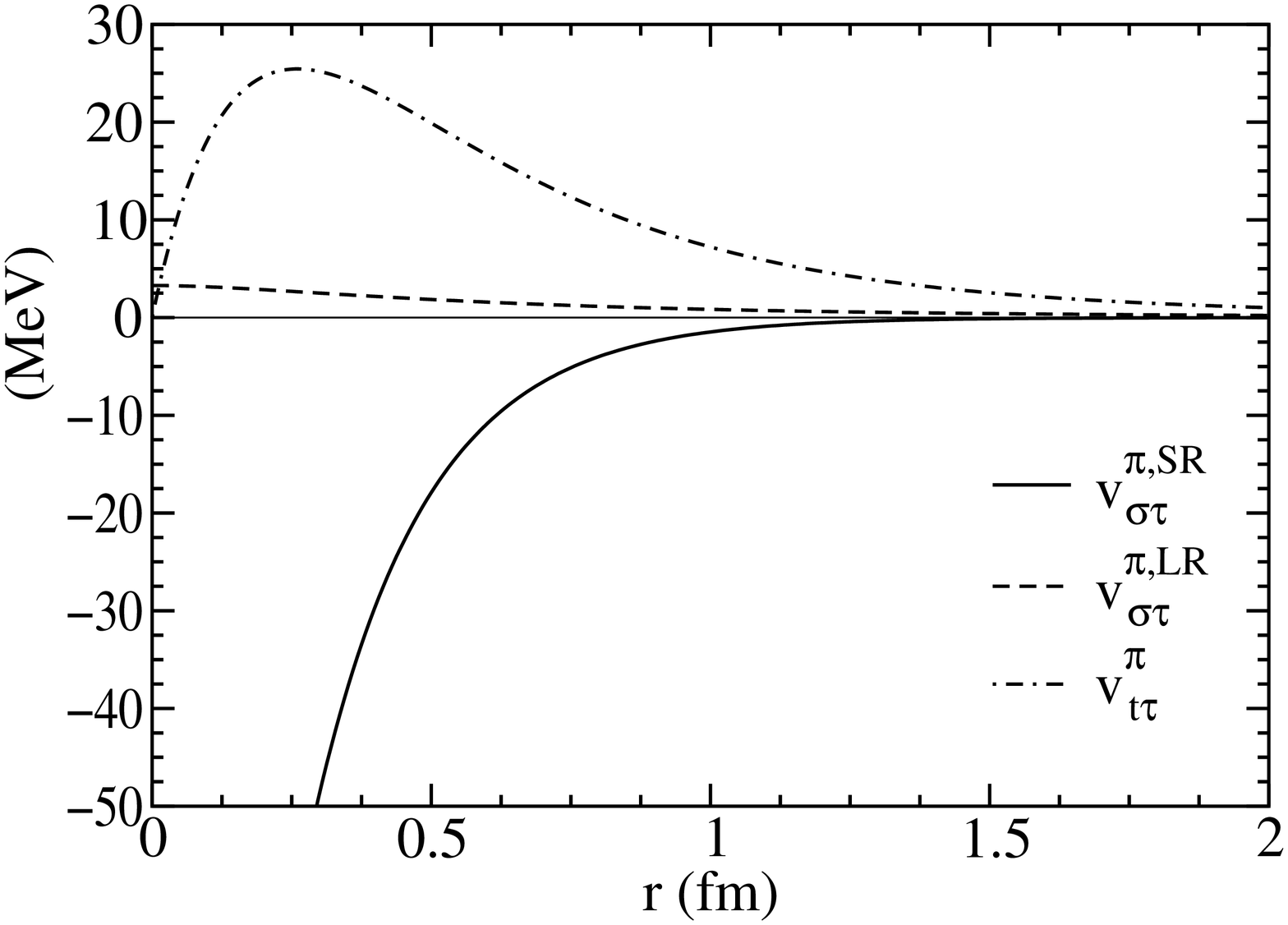}
\nopagebreak
\caption{\label{fig:opep} Regulated OPE potential in MeV.}
\end{center}
\end{figure}

The OPE interaction has the form,
\beq
\hat{v}^\pi_{ij}=
\left[
v^{\pi,SR}_{\sigma\tau}(r_{ij})+v^{\pi,LR}_{\sigma\tau}(r_{ij})
\right]
\bfsig_i\cdot\bfsig_j\bftau_i\cdot\bftau_j
+v^\pi_{t\tau}(r_{ij})S_{ij}\bftau_i\cdot\bftau_j.
\label{eqn:opep}
\eeq
For point particles, $v^{\pi,SR}_{\sigma\tau}(r_{ij})$ is the 
$\delta$-function part, while $v^{\pi,LR}_{\sigma\tau}(r_{ij})$ is 
the Yukawa function part of the $\bfsig_i\cdot\bfsig_j\bftau_i\cdot\bftau_j$ 
interaction.  The quark form factors modify the radial 
functions, plotted in Fig.(\ref{fig:opep}), as follows:
\begin{eqnarray}
\label{eqn:vpi1}
v^{\pi,SR}_{\sigma\tau}(r)&=&
{f_{\pi QQ}^2\over 4\pi}{1\over 3}\mu
\left[-{1\over 2}{\Lambda^3\over\mu^3}
\left(1-\frac{\mu^2}{\Lambda^2}\right)^2 e^{-\Lambda r}\right]\\
\label{eqn:vpi2}
v^{\pi,LR}_{\sigma\tau}(r)&=&
{f_{\pi QQ}^2\over 4\pi}{1\over 3}\mu
\left[Y_\mu(r)-{\Lambda\over\mu}Y_\Lambda(r)
-{1\over 2}\left({\Lambda\over\mu}
-{\mu\over\Lambda}\right)e^{-\Lambda r}\right]\\
\label{eqn:vpi3}
v^\pi_{t\tau}(r)&=&
{f_{\pi QQ}^2\over 4\pi}{1\over 3}\mu \nonumber \\
& &\times\left[T_\mu(r)-{\Lambda^3\over\mu^3}T_\Lambda(r)
-{1\over 2}{\Lambda\over\mu}\left({\Lambda^2\over\mu^2}-1\right)
\left(\Lambda r + 1\right) Y_\Lambda(r)\right].
\end{eqnarray}
Here $\mu=138$ MeV, the average mass of the three
charge states of the pion, and $Y_\mu$ and $T_\mu$ are as
given in Eqs.(\ref{eqn:yukawa},\ref{eqn:tensyuk}) with
$\Lambda\to\mu$. Note that $v^{\pi,SR}_{\sigma\tau}$ and
$v^{\pi,LR}_{\sigma\tau}$ have equal and opposite volume
integrals.

At large distances the sum of OPE interactions between the
nine pairs of quarks in different nucleons must be equivalent to
the $NN$ OPEP tail. Thus we fix the coupling constant,
$f_{\pi QQ}^2/4\pi$, by considering a configuration of quarks
representing two well separated nucleons. 
In this limit the ratio of the sum of OPE interactions between the quarks of 
different nucleons and the $v^{\pi}_{NN}$ contains the 
$SU(2)_{spin}\otimes SU(2)_{flavor}$ factor of $9/25$.  Therefore:
\beq
f_{\pi QQ} = {3\over5}f_{\pi NN},
\label{eqn:pioncoups}
\eeq
where the pion-nucleon coupling constant $f_{\pi NN}$ is known from
$NN$ scattering experiments to be $f_{\pi NN}^2/4\pi =0.075$\cite{stoks}.
Since the quoted value of the $\pi NN$ coupling constant is extrapolated 
to pion-pole, the monopole form factor:
\beq
\label{ff:opep}
F(\bvec{q}^2)={\Lambda^2-\mu^2 \over\Lambda^2+\bvec{q}^2},
\eeq
containing the pion mass $\mu$ is used to calculate the $\hat{v}^{\pi}$.
We assume that the cutoff parameter 
$\Lambda$ is the same for OGE and OPE interactions between quarks, 
though they could be different.

Presumably we are justified in retaining the long-range parts,
$v^{\pi,LR}_{\sigma\tau}(r)$ and $v^\pi_{t\tau}(r)$,
of OPEP. The short-range $v^{\pi,SR}_{\sigma\tau}(r)$, term
however, is the smeared delta-function
contact potential whose validity is questionable. Within the meson
exchange picture one expects an infinite number of $q \qbar$ states
to contribute at short distances. The contribution of higher mass
meson states can modify the strength of the contact term. 

The single $N$ and $\Delta$ properties calculated with this
Hamiltonian are listed in the rightmost columns of
Table \ref{tab:cfnuc}. We note several features of
these results. The kinetic energy of the quarks in the
$\Delta$--resonance is significantly
smaller than that in the nucleon, thus the $N-\Delta$ mass difference
cannot be calculated perturbatively as was noted by CKP.

\begin{figure}[hb]
\begin{center}
\includegraphics[ width= 440 pt, keepaspectratio, clip ]{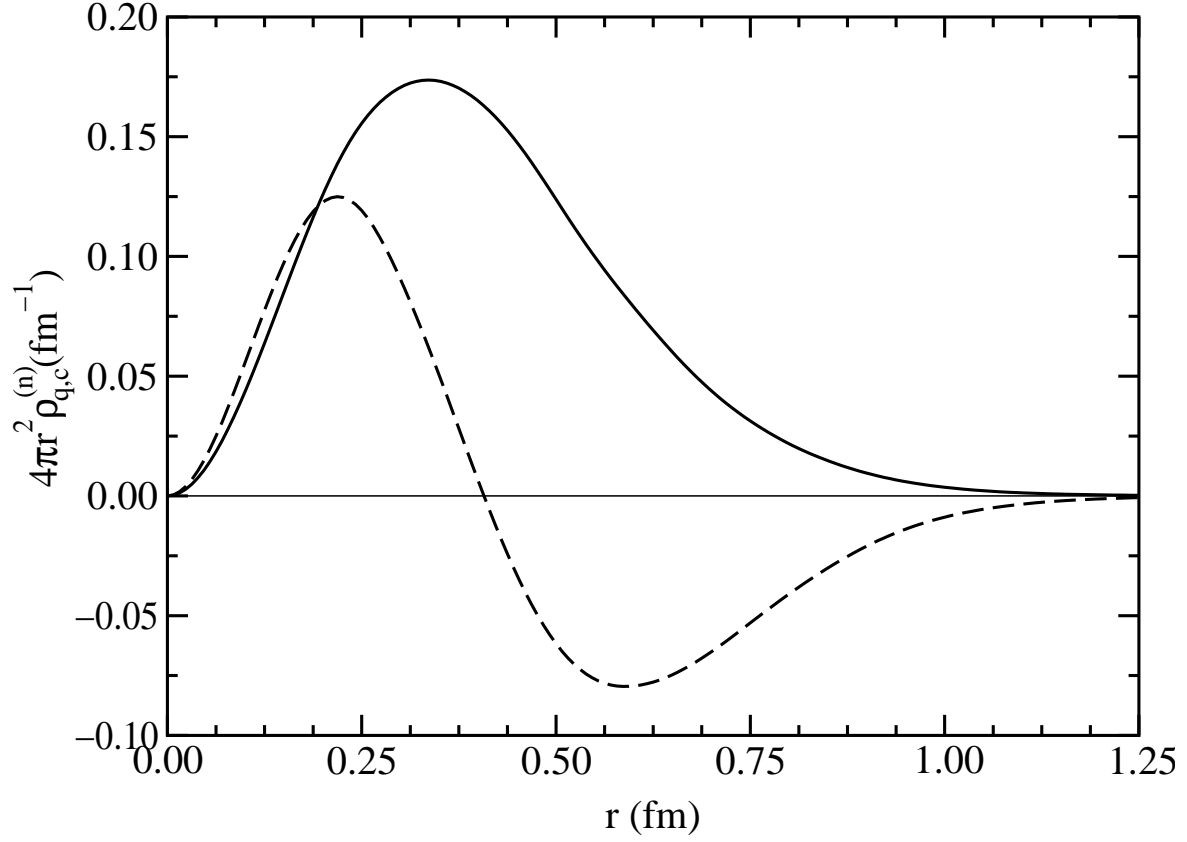}
\nopagebreak
\caption{\label{fig:rhon}
Neutron radial quark density, $4\pi r^2\rho^{(n)}_q$ (solid line)
and radial charge density, $4\pi r^2\rho^{(n)}_c$ (dashed line)
scaled by 10.}
\end{center}
\end{figure}

The $v^{\pi,SR}_{\sigma\tau}$ term makes a significant contribution
to the nucleon energy and the $N-\Delta$ mass difference. 
The nucleon is composed primarily of $T=0,S=0$ and $T=1,S=1$ pairs,
with nearly half the pairs in each channel. The value of
$\bfsig_i\cdot\bfsig_j\bftau_i\cdot\bftau_j$ in these pairs are
+9 and +1, respectively. The $\Delta$ however has $T=1,S=1$
pairs almost exclusively. This results in a strong attractive
short-range pion interaction in the nucleon relative to the $\Delta$
and causes the $d$ $(u)$ quark in the proton (neutron) to have a
smaller rms radius than the average. This effect may be seen in the
neutron charge density, $\rho^{(n)}_c$, plotted along with the quark
density $\rho^{(n)}_q$ in Fig.(\ref{fig:rhon}).

The other short-ranged interaction is the
$v^g_\sigma$ term in $\hat{v}^g_{ij}$.  Its contributions to the
energy of the nucleon and the $N-\Delta$ mass difference
are of the same order as those of $v^{\pi,SR}_{ij}$.
In fact, one can reproduce the $N-\Delta$ mass
difference with only the $v^g_\sigma$ using a larger $\alps$ and/or 
$\Lambda$. The $f_{\pi QQ}$ is fixed from the observed $f_{\pi NN}$, 
and it is necessary to have $\Lambda \sim 5$ fm$^{-1}$
to get the conventional value of $\alps$ and reproduce the
$N-\Delta$ mass difference.
The long-range spin-dependent terms, the tensor, $t\tau$, and
spin-orbit, give $\lesssim 1\%$ contribution to the total energy 
of the nucleon.  We have verified that the spectrum of
$P$-wave nucleon and $\Delta$ 
states obtained with the present Hamiltonian is as good as that 
obtained with the semi-relativistic Hamiltonian excluding OPE 
interaction used in CKP.

The {\em rms} quark radius for the nucleon is 0.44 fm.
Taking into account the size of the CQ we obtain for the
proton and neutron rms charge radii $0.67$ fm and
$-0.015$ fm, respectively.
They are smaller than the observed values of $0.79$ fm
and $-0.34$ fm.
A part of this difference could be due to the contribution of the 
pion cloud to the charge radius.

\subsection{Hamiltonian for six-quark two-nucleon states}
\label{sec:H_6q}
\begin{figure}[hb]
\input{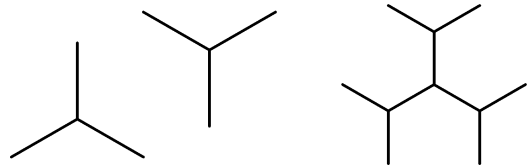}
\caption{\label{fig:flux}
Flux tube configurations for system of six
quarks consistent with gauge invariance. Fig.(a) shows
the flux tubes for the two nucleon state, while (b) shows
the ``exotic'' six-quark configuration.}
\end{figure}

Possible FT configurations for six-quark states
consistent with gauge invariance are shown in Figs.(\ref{fig:flux}a)
and (\ref{fig:flux}b). The ``exotic'' hadron configuration of
Fig.(\ref{fig:flux}b) has been shown to lie $\sim 300$ MeV above
the two nucleon state \cite{carlson:exhad} and is not considered
in this work.

Fig.(\ref{fig:flux}a) shows one of ten possible FT
configurations with two Y-junctions, corresponding to one of the 
ten ways to divide the six quarks into two indistinguishable
color-singlets listed in Table \ref{tab:partition}.
Each configuration ${\cal P} =\{1,2,\dots ,10\}$ specifies a color
state $\ket{{\cal P}}$ in which the
colors of quarks $ijk$ and $lmn$ are coupled to separate singlets.
We restrict the Hilbert space of the six-quark states to those of
the type:
\beq
\Psi = \sum_{{\cal P}=1}^{10} \Psi_{{\cal P}} \ket{{\cal P}}.
\label{psi6}
\eeq
Where the $\Psi_{{\cal P}}$ are functions of the positions, spins
and isospins of the six quarks. The $\Psi_{{\cal P}} \ket{{\cal P}}$
is antisymmetric under exchange of any pair within the singlet
$ijk$ and within $lmn$.  It is also made antisymmetric under the 
triple exchange of $ijk$ with $lmn$. The terms in the sum over ${\cal P}$
may all be reached from the wave function in a given partition,
say $\Psi_1 \ket{1}$, by quark exchange operators, $P_{ij}$ shown in
Table \ref{tab:partition}. It is useful to write Eqn.(\ref{psi6})
explicitly as,
\beq
\label{eqn:psi6p}
\Psi=\left(1-\sum_{i=1,2,3}\sum_{j=4,5,6}P_{ij}\right)\Psi_1\ket{1}.
\eeq
The minus sign in the second term ensures that this wave function
is completely antisymmetric.

Within this space, our Hamiltonian 
is a generalization of that for single nucleon discussed above.
It is given by:
\begin{eqnarray}
\hat{H} &=&
\sum_{q=1}^6 \left( m_Q - \frac{\hbar^2}{2m_Q} \nabla_q^2 \right)
+ \sum_{q < q^{\prime} \leq 6} \left( \hat{v}^g_{qq^{\prime}} 
+\hat{v}^{\pi}_{qq^{\prime}} \right) \nonumber \\
&+& \frac{1}{2} \left(
  V^C({\bf r}_i,{\bf r}_j,{\bf r}_k)
 +V^C({\bf r}_l,{\bf r}_m,{\bf r}_n) \right. \nonumber \\
&+& \left. V^C({\bf r}_{i^{\prime}},{\bf r}_{j^{\prime}},{\bf r}_{k^{\prime}})
+ V^C({\bf r}_{l^{\prime}},{\bf r}_{m^{\prime}},{\bf r}_{n^{\prime}}) \right)
- 2 V_0.
\label{eqn:H_6}
\end{eqnarray}
The quarks $ijk$ and $lmn$ are in separate color-singlets 
in the ket, $\ket{\Psi}$, while $i'j'k'$ and $l'm'n'$ are singlets
in the bra $\bra{\Psi}$.

The above Hamiltonian has ``${\cal P}$-diagonal''
or color-diagonal (CD) matrix elements, 
\beq
\label{eqn:pdiag}
H_{{\cal P}{\cal P}} = \langle \Psi_{{\cal P}} ;{\cal P} | 
\hat{H} |\Psi_{{\cal P}} ;{\cal P} \rangle ,
\eeq
in which the flux tubes remain unchanged.
In color-nondiagonal (CND) elements:
\beq
H_{{\cal P}{\cal P}^{\prime}} = \langle \Psi_{\cal P} ;{\cal P} |
\hat{H} |\Psi_{\cal P^{\prime}} ;{\cal P^{\prime}} \rangle ,
\label{eqn:pnond}
\eeq
having ${\cal P}\neq{\cal P'}$, there is an exchange of the flux tubes.
In QCD, when quarks in different nucleons are close enough 
they may exchange flux tubes.  This effect arises due to 
magnetic terms in the lattice Hamiltonian of QCD which can alter FT
paths and result in configuration mixing. The terms which
mix configurations are proportional to the inverse of the strong
coupling constant and, therefore, configuration mixing should be small
at hadronic scales where the strong coupling constant is large.

\begin{figure}[hb]
\begin{picture}(150,150)(-120,0)
\input{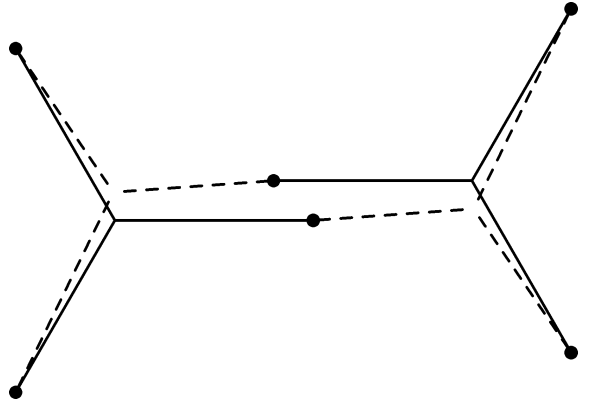}
\end{picture}
\caption{\label{fig:ftmix}
Flux tube configurations for quark exchange.
The solid lines show flux tubes for the first partition
(123;456). The dashed lines show the flux configuration
in the seventh partition (163;452), obtained by
exchanging quarks 2 and 6.}
\end{figure}

We may illustrate this point by first considering the case of
infinite strong-coupling. In this limit, for a given configuration of
quarks, the two possible FT arrangements indicated by the
sets of solid and dashed lines in Fig.(\ref{fig:ftmix}),
are orthogonal. This can be seen as follows. In the
Hamiltonian formulation of lattice QCD the gauge sector of the
theory may be written as a sum of two terms. An ``electric''
term, proportional to $g^2$, the square of the strong-coupling
constant,
and a magnetic term, which goes like $g^{-4}$. The electric
term simply counts units of flux along links of the lattice
and the magnetic term may change its path in a gauge
invariant manner. In the infinite strong-coupling limit,
$g\to\infty$, only the electric term remains.
The eigenstates of the strong-coupling Hamiltonian
are then single lines of flux along arbitrary lattice paths
which originate on quarks and terminate on Y-junctions or anti-quarks.
Thus different FT configurations are expected to be orthogonal by the
hermiticity of the strongly-coupled Hamiltonian \cite{robson:ftm}.
Due to this orthogonality the CND elements of the 
strong-coupling Hamiltonian as well as those of the normalization
will be zero.

When flux sources are close the color-magnetic terms
cause them to rearrange, resulting in FT configuration mixing.
This gives rise to CND or {\em quark exchange}
matrix elements, wherein quarks are exchanged between nucleons.
A simple way to model the supression of mixing
of significantly different FT configurations is to
insert a factor into the CND matrix elements
which falls off as the distance between the exchanged
quarks increases. We have chosen the form:
\beq
\zeta(r_{il}) = e^{-\gft^2 r_{il}^2 }
\label{eqn:ftof}
\eeq
where $\gft^{-1}$ is the range over which flux-tubes may be
exchanged, and $r_{il}$ is the distance between the exchanged
quarks $i$ and $l$. More complex parameterizations for this
factor, where $\zeta$ depends on the positions of all six quarks,
are discussed in Ref.\cite{micheal93}. The limit,
$\gft\to\infty$, corresponds to zero-range, i.e.\
no, quark exchange and is therefore referred to as
the strong-coupling limit.  The above $\zeta$ factor is
included in all CND matrix elements. We have considered
three cases: the limit $\gft\to\infty$,
finite $\gft = 2$ fm$^{-1}$, and $\gft=0$.
The $\gft=2$ fm$^{-1}$ value supresses quark
exchange for distances larger than $0.5$ fm, which is
of the order of the rms quark radius of the nucleon.

The six-quark wave functions contain the color
factors $\ket{{\cal P}}$, however, it is simple to explicitly
do the color algebra and suppress them in QMC calculations. In
the case of three-quark hadrons this corresponds to replacing all the
$\mathsf{T}_i\cdot\mathsf{T}_j$ operators by $-2/3$ and the unit
operator in color space, $\openone_C$ by one.  In the six-quark case
the color matrix elements factorize from the rest:
\beq
\bra{\Psi}\mathcal{O}(\bvec{R})\mathcal{O}^C \ket{\Psi}
 = \sum_{{\cal P'},{\cal P}}
\bra{{\cal P'}}\mathcal{O}^C\ket{{\cal P}} 
\int d\bvec{R}\: e^{-\gft^2 |\rvec_i - \rvec_l|^2}
\Psi_{{\cal P'}}^\dagger(\bvec{R})\mathcal{O}(\bvec{R})
\Psi_{{\cal P}}(\bvec{R}),
\label{eqn:matel}
\eeq
where $\mathcal{O}(\bvec{R})$ is a spin-isospin-spatial
one or two-body operator and $\mathcal{O}^C$
may be $\openone_C$ or $\mathsf{T}_q\cdot\mathsf{T}_{q'}$.
For $\mathcal{P}=\mathcal{P'}$ there is no quark exchange
and $\rvec_i - \rvec_l = 0$.  When $\mathcal{P} \neq \mathcal{P'}$ 
the exchanged quarks $i$ and $l$ 
are uniquely determined by $\mathcal{P}$ and $\mathcal{P'}$.

The ${\cal P}$-diagonal color matrix elements are trivial:
$\langle {\cal P}|\openone_C|{\cal P} \rangle = 1 $ 
and $\mathsf{T}_q\cdot\mathsf{T}_{q^{\prime}} = -2/3$ if the 
quarks $q$ and $q^{\prime}$ are in the same singlet, else 
$\mathsf{T}_q\cdot\mathsf{T}_{q^{\prime}} = 0 $. 

The CND overlap factors
\beq
\label{eqn:phasedef}
\bra{{\cal P'}}\openone_C\ket{{\cal P}}=\frac{1}{3}C_{\cal P',P}
\eeq
where the phases $C_{\mathcal{P',P}}=\pm 1$
are determined for the partitioning in Table \ref{tab:partition}
using the following rules.
The overlap of $\ket{\mathcal{P}=1}$ with any other
state $\ket{\mathcal{P'}=2,\dots,10}$ is given by,
\beq
\bb\mathcal{P'}|1\kb=\bra{1}P_{il}\ket{1} = +\frac{1}{3};
\label{eqn:lemma}
\eeq
independent of $i=1,2,3$ and $l=4,5,6$. The factor $\frac{1}{3}$
is the result for the general case of $SU(N_C)$ which is $N_C^{-1}$
with $N_C=3$. This leads to a suppression of exchange matrix
elements in comparison to colorless objects.
The remaining $C_{\mathcal{P',P}}$ may be calculated by noting that
a general overlap
$\bb\mathcal{P'}|\mathcal{P}\kb$, with $\mathcal{P'}\ne 1$
and $\mathcal{P}\ne 1$, may be written as
\beq
\bb\mathcal{P'}|\mathcal{P}\kb=\bra{1} P_{il} P_{jm} \ket{1},
\label{eqn:colol}
\eeq
where $i$ and $j$ are $\leq 3$ and $l$ and $m$ are $\geq 4$.
In the case, $i\neq j$ and $l\neq m$,
\beq
\bra{1} P_{il} P_{jm} \ket{1}=
+\frac{1}{3};
\label{eqn:case1}
\eeq
otherwise, when either $i=j$ or $l=m$,
\beq
\bra{1} P_{il} P_{jm} \ket{1}=
-\frac{1}{3}.
\label{eqn:case2}
\eeq
The last case, $i=j$ and $l=m$ is a diagonal element.

The CND ($\mathcal{P'}\neq\mathcal{P}$) factors are:
\begin{eqnarray}
\label{eqn:nondcf}
C^{(2)}_{q',q;P',P}&=&
\bra{{\cal P'}}\mathsf{T}_{q'}\cdot\mathsf{T}_q\ket{{\cal P}} \\
&=&\bra{i'j'k';l'm'n'}\mathsf{T}_{q'}\cdot\mathsf{T}_q\ket{ijk;lmn}
\end{eqnarray}
We note that
$C^{(2)}_{q',q;P',P}=C^{(2)}_{q',q;P,P'}=C^{(2)}_{q,q';P',P}$.
Using cyclicity: $i'j'k'=j'k'i'=k'i'j'$ and $l'm'n'=m'n'l'=n'l'm'$
we can always bring the above matrix element into the form:
\beq
\label{eqn:nondcf2}
C^{(2)}_{q',q;P',P}=
\bra{lj''k'';im''n''}\mathsf{T}_{q'}\cdot\mathsf{T}_q\ket{ijk;lmn}
\eeq
There are four possibilities in Eq.(\ref{eqn:nondcf2}) corresponding
to $j''=j$ or $k$ and $m''=m$ or $n$ which can differ from
\beq
\label{eqn:nondcf3}
\bra{ljk;imn}\mathsf{T}_{q'}\cdot\mathsf{T}_q\ket{ijk;lmn}
\eeq
only by a phase given by:
\beq
\label{eqn:nondcf4}
\bra{lj''k'';im''n''}\mathsf{T}_{q'}\cdot\mathsf{T}_q\ket{ijk;lmn}
=\left(2\delta_{j'',j}-1\right)\left(2\delta_{m'',m}-1\right)
\bra{ljk;imn}\mathsf{T}_{q'}\cdot\mathsf{T}_q\ket{ijk;lmn}.
\eeq
This allows us to work with just the color factor in
Eq.(\ref{eqn:nondcf3}).
There are four types of quark pairs $q',q$:
(1) For the exchanged pair $q',q=i,l$
\mbox{
$\bra{\mathcal{P'}}\mathsf{T}_{q'}\cdot\mathsf{T}_q\ket{\mathcal{P}}
= \frac{4}{9}$}.
(2) For unexchanged pairs in the same singlets, $qq'=jk$ and $mn$
\mbox{
$\bra{\mathcal{P'}}\mathsf{T}_{q'}\cdot\mathsf{T}_q\ket{\mathcal{P}}
= -\frac{2}{9}$}.
(3) For unexchanged pairs in different singlets, $qq'=jm,jn,km$ and $kn$
\mbox{
$\bra{\mathcal{P'}}\mathsf{T}_{q'}\cdot\mathsf{T}_q\ket{\mathcal{P}}
= \frac{1}{9}$}.
(4) The remaining eight pairs are between an exchanged quark
$q=i$ or $l$ and one of the unexchanged 
quarks $q'=j,k,m$ or $n$.  For these
\mbox{
$\bra{\mathcal{P'}}\mathsf{T}_{q'}\cdot\mathsf{T}_q\ket{\mathcal{P}}
= -\frac{2}{9}$}.

\section{Variational wave function}
\label{vmc}

We use the variational method of Ref.\cite{wiringa:vmc}, applied
there to light nuclei, to the problem of six interacting quarks.
The idea behind this method is that interaction
terms in the Hamiltonian will induce correlations in the wave
function which, in general, depend on the quantum numbers
of the interacting quarks. A good approximation to the ground
state eigenfunction may be obtained by applying two and three-body
correlation operators with the same operator structure as the
interaction terms in the Hamiltonian to an uncorrelated wave
function.

The six-quark wave function corresponding to two 
uncorrelated nucleons is just a product of two single-nucleon
three-quark states.  We assume that the full correlated wave function
is obtained by operating on it with correlation operators 
for pairs of quarks in different nucleons, as well as for the 
nucleon pair.

\subsection{Single nucleon wave function}
\label{sec:psi1}
The uncorrelated nucleon wave function,
$\Phi^N(m_T,m_S)$ is the product of symmetric spin-isospin
state with $T=S={1\over 2}$, $T_z=m_T$, $S_z=m_S$, 
and antisymmetric color-singlet wave functions.
The latter is eliminated from QMC calculations as described
in part (B) of the last section, and the former is, for a
spin-up proton, for example: 
\beq
\label{eqn:protuc}
\Phi^N\left(+\frac{1}{2},+\frac{1}{2}\right)={1\over 3\sqrt{2}}
\left(2\ket{u\spup u\spup d\spd}
      -\ket{u\spup u\spd  d\spup}
      -\ket{u\spd  u\spup d\spup} + \mbox{perms.}\right) .
\eeq
It has no dependence on the quark positions.
The variational nucleon wave function used in this work is,
\beq
\Psi^N(m_T,m_S) = F^I_{123}
\left( \mathcal{S} \prod_{i<j} \hat{F}^I_{ij} \right) \Phi^N(m_T,m_S).
\label{eqn:psi1}
\eeq
Here, $F^I_3$ is a three-body correlation function
and $\hat{F}^I_{ij}$ are pair correlation operators.
The superscript $I$ distinguishes correlations between quarks internal to the
nucleon. When we consider six-quark wave function, it will
have another correlation operator, $\hat{F}^E_{ij}$, which acts on
quarks in different nucleons. The symmetrized product is required
since $\hat{F}^I_{ij}$ do not commute. The $\Psi^N$ is translationally
invariant.

In CKP the CQ pair interaction contains only the confining term and
the OGE interaction. There the form of $\hat{F}^I_{ij}$ is taken to be,
\beq
\hat{F}^{(CKP)}_{ij}=[1+u_\sigma(r_{ij}) \bfsig_i\cdot\bfsig_j] f_c(r_{ij}),
\label{eqn:ckpF}
\eeq
where $f_c$ denote spatial correlations and 
$u_\sigma(r_{ij})$ is the spin-spin correlation
function induced by the $\bfsig_i\cdot\bfsig_j$ 
term in the OGEP.  The correlations induced by the tensor 
and spin-orbit parts of the OGE interaction are small, and 
neglected in CKP.

In the present work, the CQ interaction also 
includes the OPEP. We therefore consider the 
most general static spin-isospin correlation operator:
\beq
\hat{F}^I_{ij} = \left[1 + \sum_{p=2}^6 u^I_p(r_{ij}) O^p_{ij}\right]
\;f^I_c(r_{ij}),
\label{eqn:F}
\eeq
where the sum runs over the operator designations: $\tau$,
$\sigma$, $\sigma\tau$, $t$, and $t\tau$ which correspond to the
operators, numbered $p=2-6$,
\beq
O^{p=2-6}_{ij} = \bftau_i\cdot\bftau_j,\bfsig_i\cdot\bfsig_j,
\bfsig_i\cdot\bfsig_j\bftau_i\cdot\bftau_j,
S_{ij},S_{ij}\bftau_i\cdot\bftau_j \ .
\label{eqn:oplist}
\eeq
The $O^1_{ij} = \openone $ is denoted by symbol $c$.
The $\tau$ term is included for completeness, even though it
does not appear in the OGE and OPE interactions. 
We neglect the spin-orbit correlations since the spin-orbit 
interactions play a small role in the present problem.

The three-body correlation function takes into account the
small, spin-independent three-body potential of Eq.(\ref{eqn:v3}).
We take the functional form of the correlation
suggested by first order perturbation theory,
\beq
F^I_{123}=1-\beta_3 V_3(\rvec_1,\rvec_2,\rvec_3) \ ,
\label{eqn:3bF}
\eeq
where $\beta_3$ is a positive variational parameter. In fact, we
find that the value used in Ref.\cite{ckp1}, $\beta_3=0.025\times 10^{-3}$
MeV$^{-1}$ is sufficient.

The pair correlation functions, $f_c^I(r_{ij})$ and $u^I_p(r_{ij}) =
f^I_p(r_{ij})/f_c^I(r_{ij})$ are varied to minimize
the energy of the single nucleon states. Since the pair interaction 
(Eq. \ref{eqn:v2}) depends upon the total isospin and spin of the
interacting quarks it is convenient to project the correlations into 
the four possible $T,S$ channels,
\begin{eqnarray}
f^I_{T,S}(r_{ij})\ket{T,S}=
\sum_{p=1}^4 f^I_p(r_{ij})O^p_{ij}\ket{T,S} \ , \\
f^I_{t,T}(r_{ij})S_{ij}\ket{T,S}=
\sum_{p=5,6} f^I_p(r_{ij})O^p_{ij}\ket{T,S} \ .
\label{eqn:fts}
\end{eqnarray}
In a single nucleon, the antisymmetry is ensured
by the color-singlet factor, implying that the spin-isospin spatial
part of the wave function is symmetric. Thus the two-quark 
states have $T,S=(0,0)$ or $(1,1)$, for the smallest $L=0$, while
$L=1$ for $T,S=(1,0)$ or $(0,1)$.

Let $v_{T,S}^I$ and $v_{t,T}^I$ be the projections of the pair 
interaction (Eq. \ref{eqn:v2}). In absence of the third quark the 
pair correlation functions would obey the following 
two-body Schr\"{o}dinger equations with a constant $\lambda^I_{T,S}$, 
representing the eigenvalue, and $\lambda^I_{t,T} = 0$.
The equations in the $S=0$ channel are
\beq
-\hbarom [(r^{L+1}f^I_{T,0})''-\frac{L(L+1)}{r^2}(r^{L+1}f^I_{T,0})]
+ [v^I_{T,0} - \lambda^I_{T,0}](r^{L+1}f^I_{T,0})=0 ,
\label{eqn:f0}
\eeq
with $L=0,1$ for $T=0,1$ respectively.  
The coupled equations for $S=1$ are,
\begin{eqnarray}
&-\hbarom [(r^{L+1}f^I_{T,1})''-\frac{L(L+1)}{r^2}(r^{L+1}f^I_{T,1})]
+ [v^I_{T,1} - \lambda^I_{T,1}](r^{L+1}f^I_{T,1})& \nonumber \\
&+ \sqrt{8}[v^I_{t,1}-\lambda^I_{t,1}](r^{L+1}f^I_{t,T})=0 &\\
&-\hbarom [(r^{L+1}f^I_{t,T})''-\frac{6+L(L+1)}{r^2}(r^{L+1}f^I_{t,T})]
+ [v^I_{T,1} - \lambda^I_{T,1} - 2(v^I_{t,T}
- \lambda^I_{t,T})-3v^I_{\ell s,T}] (r^{L+1}f^I_{t,T}) & \nonumber \\
&+ \sqrt{8}[v^I_{t,T}-\lambda^I_{t,T}](r^{L+1}f^I_{T,1})=0&
\label{eqn:f1}
\end{eqnarray}
with $L=0,1$ for $T=1,0$.  The $L,S = 0,0;\ 1,0$ and $0,1$ equations 
are for $^1S_0$, $^1P_1$ and $^3S_1-^3D_1$ waves, while that for 
the $L,S = 1,1$ is an average for the $^3P_J$ states as discussed by 
Wiringa \cite{wiringa:vmc}.  

The effect of a third quark on the interacting pair is taken
into account by taking the $\lambda^I$ as functions parameterized as:
\begin{eqnarray}
\label{eqn:lamc}
\lambda^I_{T,S}(r)&=&\frac{\lambda^{I,0}_{T,S}}{1+e^{(r-R_{T,S})/a_{T,S}}}
+\frac{1}{6}\st r \left(1-e^{-(r/\xi_{T,S})^2}\right) \\
\label{eqn:lamt}
\lambda^I_{t,T}(r)&=&\frac{\lambda^{I,0}_{t,T}}{1+e^{(r-R_{t,T})/a_{t,T}}}.
\end{eqnarray}
The $\lambda^{I,0}$'s are adjusted to match the boundary conditions 
discussed below, and the second term in Eq.(\ref{eqn:lamc})
is required to obtain the boundary conditions. 
The parameters, $R_{T,S}$, $R_{t,T}$,
$a_{T,S}$, $a_{t,T}$, and $\xi_{T,S}$ are varied to minimize the
single-nucleon energy. We reduced this set of 16 variational parameters
to six, assuming
$R_{T,S}=R_{t,T}=R_S$, $a_{T,S}=a_{t,T}=a_S$ and $\xi_{T,S}=\xi_S$.
The small statistical variation of the total energy of the
single-nucleon state suggests that this set of six parameters
is sufficient. 

The equations are solved with boundary conditions appropriate to the
three-body system. When one quark is far from the other two,
the linear confining potential dominates in the three-body
Schr\"{o}dinger equation which reduces to,
\beq
\lim_{r\to\infty}
\left\{\frac{3\hbar^2}{4m_Q} \nabla^2(f_c^I(r))^2-\st r (f_c^I(r))^2\right\}=0.
\label{eqn:fasymp}
\eeq
The solution is the asymptotic form of the Airy function,
$r\left(f_c^{I}(r)\right)^2 \to e^{-\kappa r^{3/2}}$ with
$\kappa=16m\st /27\hbar^2$. The full set of boundary conditions
is then,
\begin{eqnarray}
f^I_{T,S}(r\to 0)&=&\mbox{constant} \\
f^I_{T,S}(r\to\infty)&=&h_{T,S}\sqrt{\frac{e^{-\kappa r^{3/2}}}{r}} \\
f^I_{t,T}(r\to 0)&=&0 \\
f^I_{t,T}(r\to\infty)&=&\eta_T f^I_{T,1}(r)
\label{eqn:bc}
\end{eqnarray}
where the central normalizations, $h_{T,S}$, and $D/S$ ratios,
$\eta_T$, are further variational parameters. Only three
normalization factors, $h_{T,S}$, are independent
since one sets the overall
normalization of the wave function.

\begin{figure}[hb]
\begin{center}
\includegraphics[ width= 340 pt, keepaspectratio, clip ]{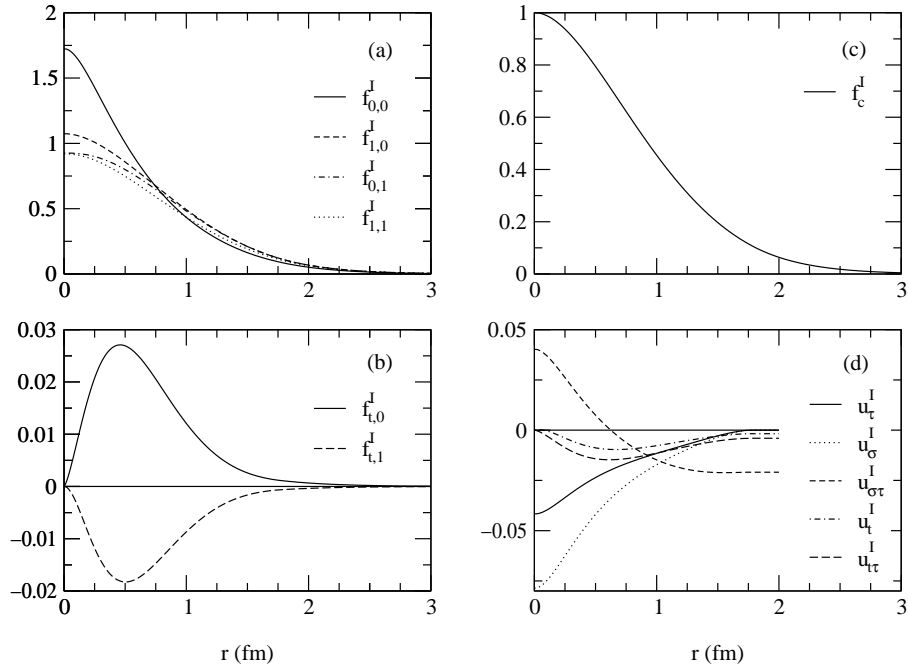}
\nopagebreak
\caption{\label{fig:fint}
Internal pair correlation functions in nucleon:
(a) central, projected into $T,S$ channels, $f^I_{T,S}$;
(b) $S=1$ tensor, $f^I_{t,T}$; (c) central correlation,
$f^I_c$; (d) operator correlations, $u^I_p=f^I_p/f^I_c$,
at $r<2$. They are constant for $r>2$ fm.}
\end{center}
\end{figure}

The optimum internal correlations are plotted in Fig.(\ref{fig:fint}).
At small distances, $\lesssim0.5$ fm the
spin-isospin dependence is seen in the relative sizes of the
$f^c_{T,S}$. We notice that $T,S=0,0$ pairs are favored
at small distance due to the attractive short-range
terms in the OGE and OPE interactions. 
This results in the $u$-quark ($d$-quark) distributions in the
neutron (proton) having smaller radii than their isospin partner
and gives a negative rms charge radius to the neutron, as mentioned
earlier. 

\subsection{Two-Nucleon variational wave function}
\label{sec:corrwf}
The six-quark wave function representing two uncorrelated nucleons 
in $T,S=0,1$ state with spin projection $M_S$ is given by:
\begin{eqnarray}
\label{eqn:PhiNN}
\Phi^{NN}(M_S)
&=& \sum_{{\cal P}=1}^{10} (-1)^{(1+\delta_{{\cal P},1})} 
\Phi^{NN}_{\cal P}(M_S) \ket{\mathcal{P}} \\
\label{eqn:PhiU}
\Phi^{NN}_{\cal P}(M_S) &=& \frac{1}{\sqrt{2}} \left[
\Psi^N_{\cal P}(p,m_1;ijk) \Psi^N_{\cal P}(n,m_2;lmn)\right. \nonumber \\
&  & - \left. \ \ \ \Psi^N_{\cal P}(n,m_1;ijk) \Psi^N_{\cal P}(p,m_2;lmn)
\right]^{S=1}_{M_S}
\end{eqnarray}
Here $\Psi^N_{\cal P}(p(n),m_1;ijk)$ denotes the wave function of
quarks $ijk$ in the proton (neutron) with
spin projection $m_1$, as in Eq.(\ref{eqn:psi1}).
The quark numbers $ijk$ and $lmn$ depend upon the partition 
${\cal P}$, and the full wave function $\Phi^{NN}(M_S)$ is 
antisymmetric under the exchange of any quark pair.  
The spins inside square brackets are coupled to values indicated 
by the sub- and superscripts.

The variational six-quark wave function, $\Psi_V$
for the interacting two-nucleon
state is obtained by inserting correlations in the above wave function. 
We consider long range correlations between the centers-of-mass of
quarks $ijk$ and $lmn$, as well as correlations between the nine 
external quark pairs, like $il$, with one quark from each singlet.  
The correlations between internal quark pairs, such as $ij$ in the 
same singlet, are included in the nucleon 
wave functions $\Psi^N_{\cal P}(p(n),m_1;ijk)$.  The $\Psi_V$ is
chosen as:
\begin{eqnarray}
\label{eqn:Psi6V}
\Psi_V(M) &=& \sum_{\mathcal{P}=1}^{10} (-1)^{(1+\delta_{{\cal P},1})}
\left(\mathcal{S} \prod_{q=i,j,k \atop q'=l,m,n}
\hat{F}^E_{qq'} \right)
\Psi^{NN}_{\cal P}(M)\ket{\mathcal{P}} \ ,\\
\label{eqn:PsiNN}
\Psi^{NN}_{\cal P}(M) &=&
\frac{\tilde{u}(r_{\cal P})}{r_{\cal P}} Y_{0}^0 \Phi^{NN}_{\cal P}(M_S=M)
+\frac{\tilde{w}(r_{\cal P})}{r_{\cal P}}\left[ Y_{2}^{M_L}
(\hat{\rvec}_{\cal P})\Phi^{NN}_{\cal P}(M_S) \right]^{J=1}_{M_J=M}
\end{eqnarray}
where
\beq
\label{eqn:relcom}
{\bf r}_{\cal P} = \frac{1}{3}({\bf r}_i+{\bf r}_j+{\bf r}_k
-{\bf r}_l-{\bf r}_m-{\bf r}_n ).
\eeq
The long range correlations, $\tilde{u}(r_{\cal P})$
and $\tilde{w}(r_{\cal P})$,
are calculated from the two-nucleon Schr\"{o}dinger equation,
\beq
\label{eqn:NNse}
\left\{-\frac{\hbar^2}{m_N}\nabla^2 + \vnn(\rvec) \right\}\Psi^{NN}(\rvec)
= E^{NN} \Psi^{NN}(\rvec).
\eeq
Here $\Psi^{NN}$ is the wave function of Eq.(\ref{eqn:PsiNN}) taken
in a given partition, say the first. The substitution of $\Psi^{NN}$
into Eq.(\ref{eqn:NNse}) results in the set of coupled equations
for $\tilde{u}$ and $\tilde{w}$ shown in
Eqs.(\ref{eqn:urel},\ref{eqn:wrel}) with
$u\to\tilde{u}$ and $w\to\tilde{w}$.
These are solved subject to the boundary conditions:
\begin{eqnarray}
\label{eqn:relbc}
\tilde{u}(R_c)=0. \nonumber \\
\tilde{w}(R_c)=0.
\end{eqnarray}
for the ground state in a cavity of diameter $R_c$.
The two-nucleon effective potential $\vnn$, used to
minimize the energy, is discussed in the next section.

The external correlation operators, $\hat{F}^E_{qq'}$, contain six
terms associated with the operators $O^{p=1,6}_{qq'}$, as do the 
the internal correlations.  By projecting the $\hat{F}^E_{qq'}$ 
into two-quark $T,S$ channels we obtain the six functions $f^E_{T,S}$
and $f^E_{t,T}$. These 
functions are assumed to obey Eqs.(\ref{eqn:f0}--\ref{eqn:f1}) with the
internal $v^I$ and $\lambda^I$ replaced by the external $v^E$ and
$\lambda^E$, however the boundary conditions of the $f^E$ are
different from those of $f^I$ as discussed below.

Terms without quark exchange between the nucleons, i.e.\ 
those having $\mathcal{P}=\mathcal{P'}$ seem to dominate the 
expectation values of $\openone$ and $\hat{H}$.  In these terms 
the confinement as well as the OGE interactions do not 
contribute to external pairs.  Therefore we take:
\beq
\hat{v}^E_{ij} = \hat{v}^\pi_{ij}(\rvec_{ij}).
\label{eqn:vext}
\eeq

The boundary conditions for the external correlations are those
appropriate for objects which are free at large separation. They are:
\begin{eqnarray}
\label{eqn:ebc}
f^E_{T,S}(r\to 0)&=&\mbox{constant} \\
f^E_{T,S}(r\ge d_c)&=&1 \\
f^E_{t,T}(r\to 0)&=&0 \\
f^E_{t,T}(r\ge d_t)&=&0
\end{eqnarray}
where $d_c$ and $d_t$ are the central and tensor ``healing'' lengths
which are varied to minimize the variational energy.

\begin{figure}[hb]
\begin{center}
\includegraphics[ width= 340 pt, keepaspectratio, clip ]{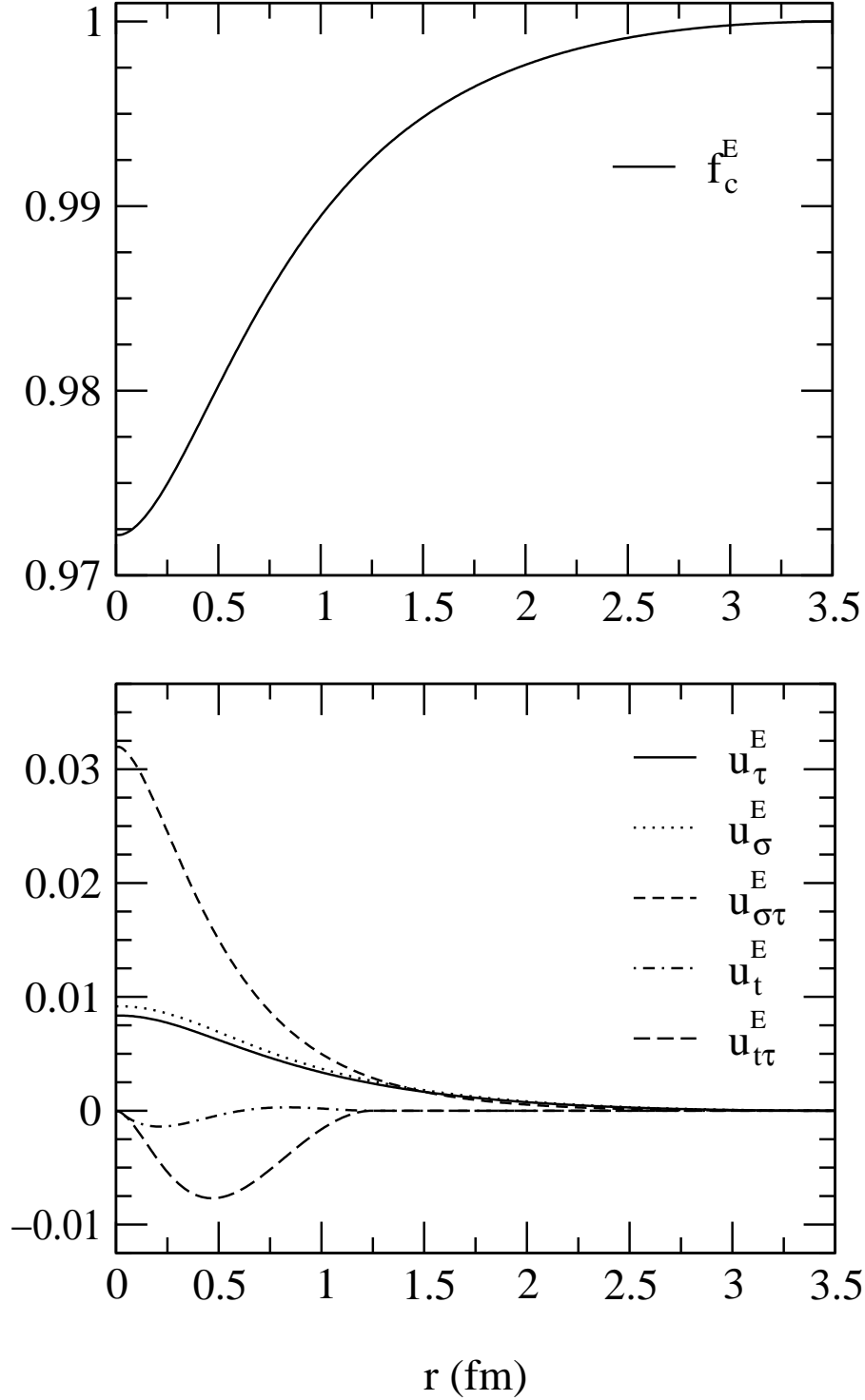}
\nopagebreak
\caption{\label{fig:fext}
External pair correlation functions in six-quark
wave function for $R_c=2$ fm with $d_c=3.55, d_t=1.25$:
central correlation, $f^E_c$, (top); operator correlations,
$f^E_p$ (bottom).}
\end{center}
\end{figure}

As in the theory of nuclear matter \cite{wiringa:nm} the
$\lambda^E$ are taken as constants for distances less
than the healing lengths while the boundary conditions
fix the form for radii greater than the healing lengths:
\begin{eqnarray}
\lambda^E_{T,0}(r)&=&\lambda^{E,0}_{T,0}\theta(d_c-r)+v_{T,0}\theta(r-d_c) \\
\lambda^E_{T,1}(r)&=&\lambda^{E,0}_{T,1}\theta(d_c-r)
+[v^E_{T,1}+\sqrt{8}(v^E_{t,T}-\lambda^E_{T,1})f_{t,T}]\theta(r-d_c) \\
\lambda^E_{t,T}(r)&=&\lambda^{E,0}_{t,T}\theta(d_t-r)+v_{t,T}\theta(r-d_t),
\label{eqn:extlam}
\end{eqnarray}
with $\theta(r)$ the Heaviside step function.  The $\lambda^{E,0}$ 
are varied to match the boundary conditions, so there are no
free parameters besides $d_c$ and $d_t$.
The external correlations are shown in Fig.(\ref{fig:fext})
for the optimum values of $d_c,d_t$ at $R_c=2$ fm.
The central correlation, $f^E_c(r)$ is seen to be $\simeq 1$ over
the entire range, with a slight `dip' at short distances which
arises primarily due to the short-ranged $\sigma\tau$ term in OPEP.
Even though the $f^E$ are small, they have an effect since the total
wave function has a symmeterized product of nine of them.

\subsection{Monte Carlo evaluation}
\label{sec:monte}
The present VMC calculation of six-quark states is formally
similar to that of $^6$Li \cite{pudliner1} except
for the color factors. The wave function
Eq.(\ref{eqn:Psi6V}) is expressed as a vector function 
of the configuration vector ${\bf R}=\{{\bf r}_1,...{\bf r}_6\}$: 
\begin{eqnarray}
\Psi_V &=&\sum_{\mathcal{P}=1}^{10} \Psi_{\cal P}({\bf R})\ket{\mathcal{P}},
       \nonumber \\
       &=&\sum_{\mathcal{P}=1}^{10} \sum_{\alpha = 1}^{320}
       \psi_{{\cal P},\alpha}({\bf R})\ket{\alpha}\ket{\mathcal{P}},
\label{eqn:PsiVMC} 
\end{eqnarray}
where $|\alpha \rangle $ are six particle spin-isospin states.  
The tensor correlations populate all the $2^6$ spin states,  
and, since isospin is conserved, we consider only the 
five possible $T=0$ six-quark states.
The set of the 320 states $|\alpha \rangle $ 
used here is the same as that in $^6$Li.
In this representation the correlation operators $\hat{F}_{qq^{\prime}}$ 
are sparse, $320\times 320$ matrix functions of ${\bf R}$.
The wave function of Eq.(\ref{eqn:Psi6V}) contains symmetrized products
of correlation operators $\hat{F}^I_{qq^{\prime}}$ and
$\hat{F}^E_{qq^{\prime}}$. Each $\Psi_{\cal P}({\bf R})$ is a sum
of $N_{pr}!$ terms, where the number of pairs, $N_{pr}=15$.
The $\Psi_{\mathcal{P},r}$ denotes the term in which the $N_{pr}$
pair correlation operators act in an order denoted by $r$, and
\beq
\Psi_{\cal P}({\bf R}) = \sum_{r=1}^{N_{pr}!} \Psi_{{\cal P},r}({\bf R}).
\label{eqn:PsiVMCs} 
\eeq
This sum is evaluated stochastically in Monte Carlo.

The expectation value of an operator: 
\beq
\label{eqn:expval}
\bb\hat{O}\kb = \frac{\bra{\Psi_V}\hat{O}\ket{\Psi_V}}{\bb\Psi_V|\Psi_V\kb}
              = \frac{\cal N}{\cal D},
\eeq
is evaluated using $N_c$ samples of ${\bf R}_I$, and left, right 
operator orders $l_I$ and $r_I$ obtained by sampling the weight:
\beq
\label{eqn:weight}
W_{l,r}(\bvec{R})=| \sum_{\mathcal{P,P'}}
  (-1)^{\delta_{{\cal P},1}+\delta_{{\cal P'},1}}
  \Psi_{{\cal P},l}^\dagger(\bvec{R})
  \Psi_{{\cal P'},r}(\bvec{R})
  \bb \mathcal{P} | \mathcal{P}' \kb
  e^{-\gamma_{FT}^2|\rvec_i-\rvec_j|^2}|,
\eeq
The average value and the standard deviation are calculated from
$\sim 10^5$ samples using block averaging.
The CD and CND contributions are evaluated separately; for example
${\cal N} = {\cal N}_{CD} + {\cal N}_{CND}$, where ${\cal N}_{CD}$
is the average value of:
\beq
\sum_{\cal P} \Psi_{{\cal P},l_I}^\dagger(\bvec{R}_I)
   \hat{O} \Psi_{{\cal P},r_I}(\bvec{R}_I)\ ,
\label{eqn:cd}
\eeq
while ${\cal N}_{CND}$ contains all the other terms.

The Fortran{\bf 90} computer code for this calculation was run
on 6 nodes of an IBM-SP . The amount of computer time required to
generate one statitistically independent configuration
$\{\bvec{R},l,r\}$ and evaluate its energy is $1.35$ node seconds.
For runs of $10^5$ configurations, we
obtain a statistical variance for the total energy of the
system at the $\sim 0.1$\% level. A run of this length takes
$\sim 40$ node hours.

\section{Six-body results}
\label{sec:results}
The parameters for the single particle Hamiltonian, 
Eq.(\ref{eqn:H_N}), as previously described, are 
listed in Table \ref{tab:hpars}.
As mentioned earlier the $\alps$ is determined from the
$N-\Delta$ mass difference and $V_0$ is chosen to reproduce the
nucleon mass, $m_N=938.9$ MeV, taken as the average of the
proton and neutron masses. We choose two values for the flux-tube
overlap parameter, \gft, to cover the physically interesting range,
and minimize the energy of $\Psi_V$ with respect to variational
parameters for each of them. The energy is also calculated
for the $R_c=2$ fm case, with $\gft=0$, perturbatively, as
discussed below.

The six-body wave function is minimized
with respect to the external pair correlation operators,
$\hat{F}^E_{ij}$, and the $NN$ long range correlation functions,
$\tilde{u}$ and $\tilde{w}$. The $\hat{F}^E_{ij}$ are varied by
adjusting the central and tensor healing lengths, $d_c$ and $d_t$.
Internal pair correlation operators, $\hat{F}^I_{ij}$,
as well as the internal triplet correlation function, $F^I_{123}$,
are held at the equilibrium values for the single nucleon.
The $\tilde{u}$ and $\tilde{w}$ in Eq.(\ref{eqn:PsiNN}),
are obtained from Eq.(\ref{eqn:NNse})
with an effective $NN$ interaction, \vnn\ and subject to the boundary
conditions, Eqs.(\ref{eqn:relbc}).
We obtain an approximate effective potential, $\hat{v}_{con}$
by the ``convolution'' method:
\beq
\label{eqn:conv}
v_{con}(R;M)=\int d\bvec{R}\ 
\delta^{(3)}\!\!\left(\bvec{R}_1-\frac{R}{2}\bvec{\hat{z}}\right)
\delta^{(3)}\!\!\left(\bvec{R}_2+\frac{R}{2}\bvec{\hat{z}}\right)
\Phi^{NN\dagger}_1(M)
\left(\,\sum_{q=1,2,3}\sum_{q'=4,5,6} \hat{v}^{\pi}_{qq'}\right)
\Phi^{NN}_1(M)
\eeq
with $d\bvec{R}=d^3r_1 d^3r_2\cdots d^3r_6$,
$\bvec{R}_1=\frac{1}{3}(\rvec_1+\rvec_2+\rvec_3)$,
$\bvec{R}_2=\frac{1}{3}(\rvec_4+\rvec_5+\rvec_6)$,
and $\Phi^{NN}_1$ given by Eq.(\ref{eqn:PhiU}) with
$\mathcal{P}=1$. The delta-functions
fix the centers-of-mass of the nucleons at separation $R$ along
the $z$-axis. Note that $\Phi^{NN}_1$ depends only on the internal
quark pairs in the first partition $\mathcal{P}=1$ and is independent
of the nucleon centers-of-mass, $\bvec{R}_1$ and $\bvec{R}_2$.
Only the OPEP between quarks in different nucleons contributes
to this convolution potential; it therefore has no spin-orbit term.
We evaluate $v_{con}(R;M)$ via Monte Carlo integration for two
values of $M=1,0$ and project into central and tensor
potentials, $v_{con}^c$ and $v_{con}^t$ as:
\begin{eqnarray}
\label{eqn:vnnc}
v_{con}^c(R) &=& \frac{1}{3}\left[2v_{con}(R;1)+v_{con}(R;0)\right] \\
\label{eqn:vnnt}
v_{con}^t(R) &=& \frac{1}{6}\left[ v_{con}(R;1)-v_{con}(R;0)\right].
\end{eqnarray}
These are plotted in Fig.(\ref{fig:vnn}) along with the
corresponding terms from the Argonne \argonne\ potential which
appear in Eqs.(\ref{eqn:urel},\ref{eqn:wrel}).

\begin{figure}[hb]
\begin{center}
\includegraphics[ width= 375 pt, keepaspectratio, clip ]{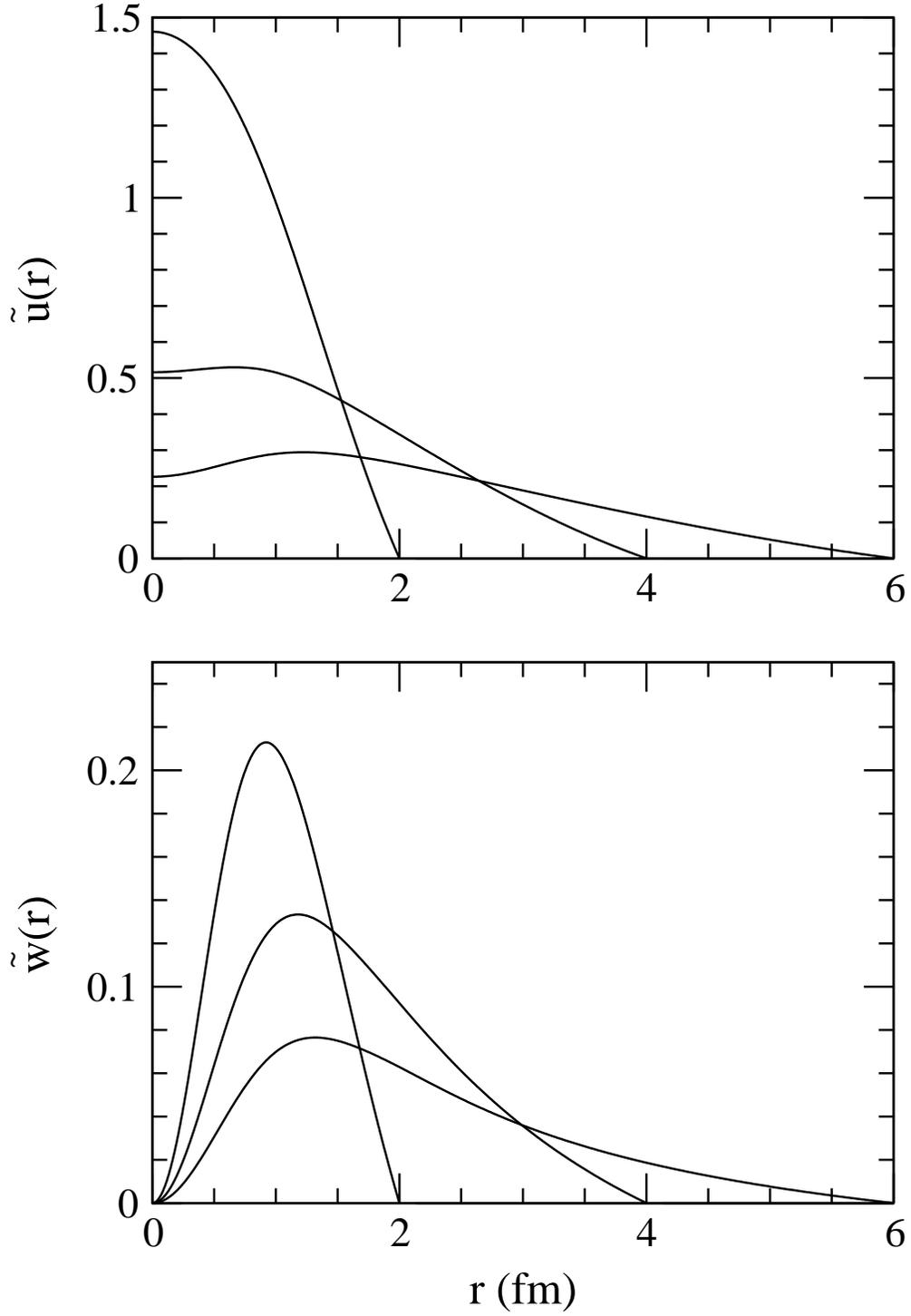}
\nopagebreak
\caption{\label{fig:effwf} $NN$ radial correlation functions $\tilde{u}(r)$
(upper part) and $\tilde{w}(r)$ (lower part) for $R_c=2$, 4, and 6 fm.}
\end{center}
\end{figure}

The effective potential appearing in Eq.(\ref{eqn:NNse}) is taken as:
\beq
\label{eqn:varvnn}
\hat{v}_{NN}=\alpha_c v_{con}^c+\alpha_t v_{con}^t S_{12}.
\eeq
where $\alpha_c$ and $\alpha_t$ are determined variationally
for each $R_c$ and \gft.  The resulting $\tilde{u}$ and $\tilde{w}$
for $\gft=2$ fm$^{-1}$ are plotted in Fig.(\ref{fig:effwf}).
They should not be interpreted as the $NN$ two nucleon relative
wave functions since the six-quark $\Psi_V$ contains additional
$f^E$ correlations. The $NN$ radial wave functions may be
obtained from $\Psi_V$ and $\Psi^{NN}$, as discussed in
Sec.\ref{sec:concl}. Table \ref{tab:params} gives the optimum values
of $d_c$, $d_t$, $\alpha_c$, and $\alpha_t$. As discussed later,
our $\Psi_V$ is less accurate for $R_c=6$ fm, $\gft\to\infty$ than
for all other cases. The $\alpha_c$ and $\alpha_t$ for this case
may not be the optimum values.

Smaller values of $\alpha_c$ are favored presumably because the
external pair correlation function, $f^E_c(r_{ij})$, less
than one for small distances, supresses configurations where
the quarks in different nucleons approach each other.
When the nucleons overlap the small difference in $f^E_c(r_{ij})$
from unity has a significant effect on the wave function since it
contributes via all the nine pairs of quarks belonging to different
nucleons. The facts, that $\alpha_t\approx 1$ in all states and
$d_t$ is small, suggest that most of the tensor two-nucleon correlation
is induced by the effective \vnn.

The results of the full VMC calculation are given in
Table \ref{tab:cfnuc} for each $R_c$ and \gft.
The top three rows of this table give the total, kinetic,
and potential energies for the optimum $\Psi_V$ per nucleon.
The statistical variance relative to the total energy is:
\beq
\label{eqn:var}
\sigma_{rel} = \left(
	   \frac{N_c^{-1}\sum_I E_V^2(\bvec{R}_I)-\overline{E_V}^2}
	   {\overline{E_V}^2}\right)^{1/2} = \sqrt{\overline{\epsilon^2}}
\eeq
where $E_V(\bvec{R}_I)$ is the local energy, Eq.(\ref{eqn:localen}),
$\overline{E_V}$ is the average of the local energy, and $N_c$ is
the number of configurations, $\bvec{R}_I$.
The relative fluctuation, $\epsilon(\bvec{R}_I)$, in $E_V(\bvec{R}_I)$
is defined as $E_V(\bvec{R}_I)=\overline{E_V}(1+\epsilon(\bvec{R}_I))$,
and $\sigma_{rel}$ gives its rms value.
If $\Psi_V$ is the true ground state wave function then
$E_V(\bvec{R}_I)=\overline{E_V}$, for all $\bvec{R}_I$ and
$\epsilon=0$. The rms value of $\epsilon$ is $0.13$
in all states in Table \ref{tab:cfnuc} except
$\gft\to\infty$, $R_c=6$ fm where $\sigma_{rel}=0.17$.
The perturbation calculation, $R_c=2$ fm, $\gft=0$ has
$\sigma_{rel}=0.19$. We can probably reduce these $\epsilon$
values by reoptimizing $\alpha_c$, $\alpha_t$, $d_c$, and $d_t$,
but the present results serve to indicate the role of the external
$f^E$ and $\tilde{u}$ and $\tilde{w}$ in improving $\Psi_V$.
The fluctuations in the kinetic and
total potential energies, $T$ and $V$, relative to the total
energy, are much larger, of $\sim\!1$. Although $T$ and $V$
fluctuate wildly as $\bvec{R}_I$ varies, their sum,
$E_V(\bvec{R}_I)$, changes little, evidence that $\Psi_V$
is a good representation of the true ground state, $\Psi_0$.

The standard deviation in the expectation values obtained from
$10^5$ configurations is shown in parentheses for all
observables in Table \ref{tab:cfnuc}, and the deviation
in $E_V$ is only 0.4 MeV per nucleon. This table
also lists the contributions of the various potential terms
in the Hamiltonian, and the energy of $\Psi_V$
relative to two free nucleons, per nucleon, compared to the
empirical value calculated from Argonne \argonne. It
shows that all of the terms in the Hamiltonian differ by only a
few percent from their single nucleon values, except 
$v^\pi_{t\tau}$, which is greatly enhanced. Comparison of the
calculated $\delta E(R_c)$ with $\delta E_{emp}(R_c)$ shows that the
present six-quark model does not give a sufficiently
attractive strong interaction. The quark kinetic energy gives
a repulsive contribution to $E(R_c)$, while the interaction
terms give a much smaller attraction. For example, at $R_c=2$ fm
the $NN$ values for the total kinetic and potential energies
per nucleon in Table \ref{tab:av18} are 80 and $-43$ MeV,
respectively, while the six-quark values are $62$ and $-3$,
relative to two free nucleons. The kinetic energy is of the
expected size but the potential is not attractive enough.

Comparison of $\gft=2$ fm$^{-1}$ and $\infty$ results indicates
that the total effect of CND contributions, via flux-tube or
quark exchange, forbidden for $\gft\to\infty$, is small and
repulsive. A perturbative calculation with the $\gft=2$ fm$^{-1}$
wave function for $\gft=0$, i.e.\ no exchange suppression due to 
flux-tube overlap, gives an even higher energy of
$1015.0(6)$ MeV/nucleon at $R_c=2$ fm, corresponding to
$E(R_c)=76.1(7)$ MeV/nucleon and $\delta E(R_c) = 24.9(7)$ MeV/nucleon.
We breakdown the expectation values of operators into CD and
CND terms, listed in Tables \ref{tab:rc2},\ref{tab:rc4} and
\ref{tab:rc6}, according to:
\beq
\label{eqn:dvx}
\bb \hat{O} \kb = \frac{\mathcal{N}_{CD}+\mathcal{N}_{CND}}
{\mathcal{D}_{CD}+\mathcal{D}_{CND}}
\eeq
where the terms in the numerator and denominator are evaluated
as in Sec.\ref{sec:monte}. The CND term of the unit
operator, $\mathcal{D}_{CND}$, is small so we expand the
denominator of Eq.(\ref{eqn:dvx}) and obtain the CD contribution
as $\mathcal{N}_{CD}/\mathcal{D}_{CD}$, while
the CND part of the expectation value is:
\beq
\label{eqn:cnd}
 \frac{\mathcal{N}_{CND}}{\mathcal{D}_{CD}}
-\frac{\mathcal{D}_{CND}}{\mathcal{D}_{CD}}
 \frac{\mathcal{N}_{CD}} {\mathcal{D}_{CD}}.
\eeq
As can be seen from Tables \ref{tab:rc2} to \ref{tab:rc6}, the
variance of the CD and CND contributions to the energy is larger
than that of the total.

For short range quark exchange, $\gft=2$ fm$^{-1}$,
the CND contribution of the unit operator, $\mathcal{D}_{CND}$,
is $-0.0142$, $-0.0017$, and $-0.0003$ with the wave function
normalized to one, for $R_c=2$, 4, and 6 fm, respectively.
However, the sign of the $\mathcal{D}_{CND}$ changes for long range
quark exchange, $\gft=0$, $R_c=2$ fm, having the value $+0.0058$.
It is simple to understand why the exchange contribution is expected
to be small and positive for long range quark exchange, so we
consider this case first. If we ignore all spatial correlation of
quarks in the wave function $\Phi^{NN}(M_S)$ of Eq.(\ref{eqn:PhiNN}),
we obtain a CND contribution of $+1/9$ to the normalization.
Localization of the nucleons reduces this value by suppression
of exchanges for separations of the exchanged pair larger than
the single nucleon rms quark diameter.

\begin{figure}[hb]
\begin{center}
\includegraphics[ width= 375 pt, keepaspectratio, clip ]{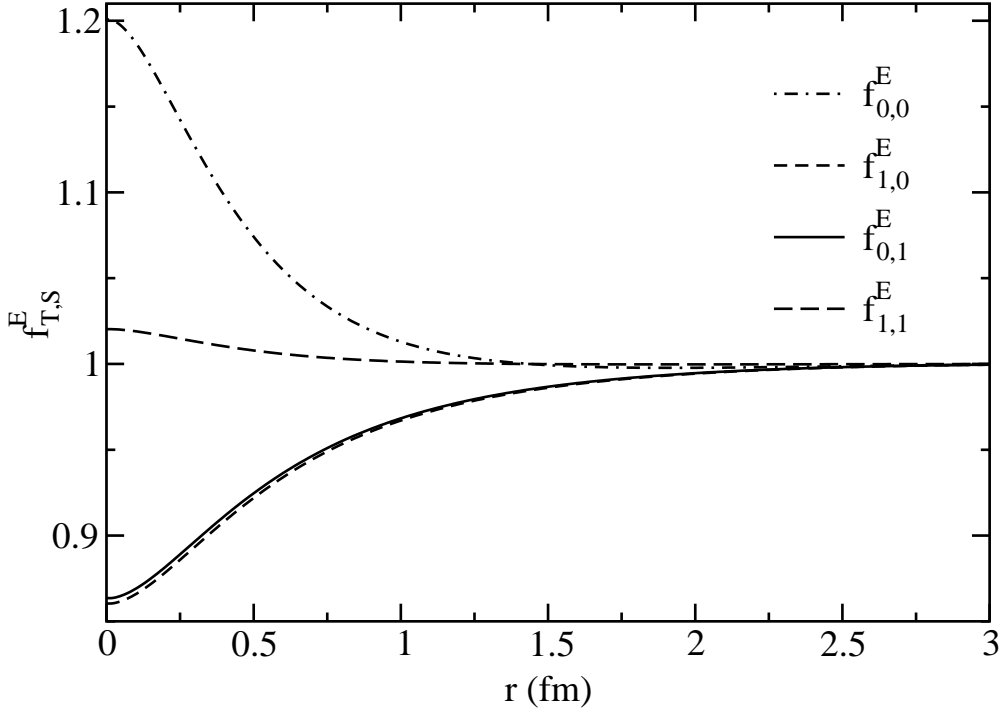}
\nopagebreak
\caption{\label{fig:fextts} External central correlation operators
projected into $T,S$ channels.}
\end{center}
\end{figure}

In the case of $\gft=2$ fm$^{-1}$ the sign of the CND contribution to
the normalization is a consequence of the spin-isospin-color dependence
of the wave function. Overall antisymmetry of the wave function implies
that the sign of the contribution for exchanged quark pairs in the
$T,S=(0,0)$ and $(1,1)$ states is negative while those in $(1,0)$ and
$(0,1)$ are positive. External quark pairs interact via OPEP, cf.\ 
Eq.(\ref{eqn:vext}), resulting in the strong spin-isospin dependence
of the external pair correlation functions, $f^E_{T,S}$, shown in
Fig.(\ref{fig:fextts}). The $\hat{v}^\pi_{ij}$ is attractive at short
range in $T,S=(0,0)$ and $(1,1)$ states resulting in the enhancement
of exchanges which make negative contributions to the CND normalization.
Similar reasoning shows that the exchange contribution due to 
$T,S=(1,0)$ and $(0,1)$ pairs is suppressed.

Though the quark exchange contributions are
$<1\%$ of $E_V$, they are $\sim 10\%$ of the $E(R_c)$, which
is related to the two-nucleon interaction, and add to its repulsive
core. They significantly effect the total effect of strong
interactions, being the same order as $\delta E(R_c)$.

\section{Conclusions}
\label{sec:concl}
We have shown that it is possible to obtain a variational
wave function for the six-quark system which gives
a small standard deviation in the energy for relatively
modest amounts of computational time. The present study also
shows that exact calculations with small statistical errors are
possible with this Hamiltonian. A GFMC code is currently
being developed to project out the true ground state component
of $\Psi_V$ using Euclidean propagation, 
$\Psi_0 = \lim_{\tau\to\infty} e^{-(\hat{H}-E_0)\tau} \Psi_V$.

\begin{figure}[hb]
\begin{center}
\includegraphics[ width= 400 pt, keepaspectratio, clip ]{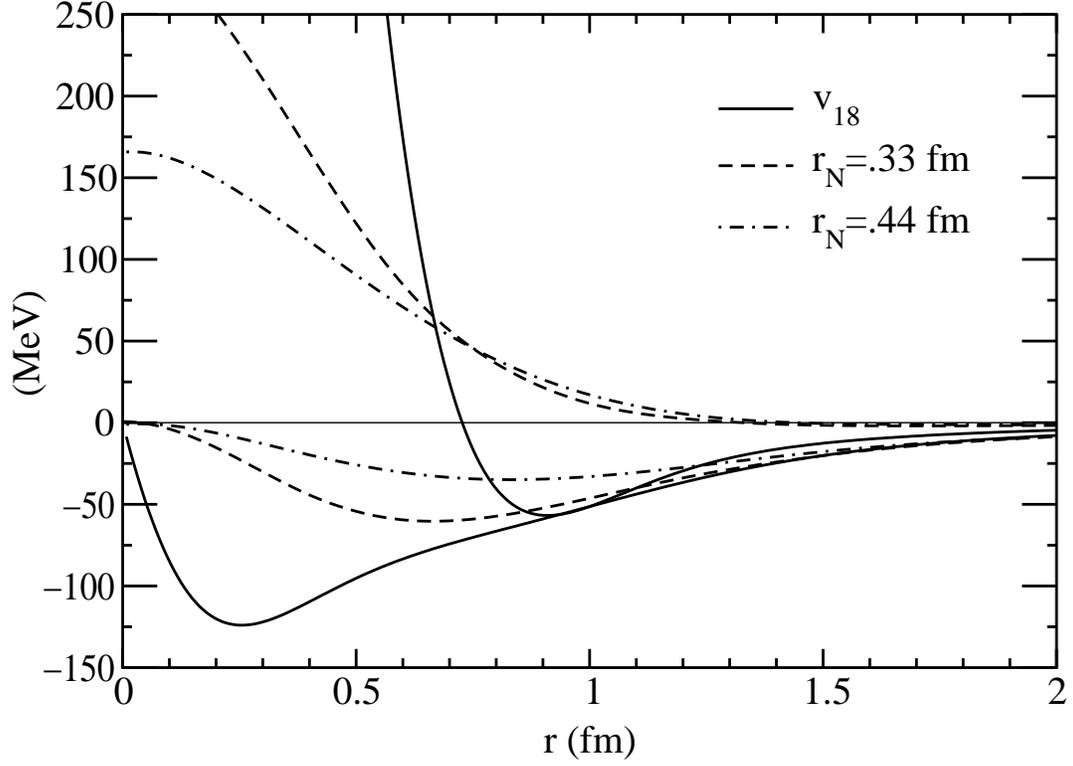}
\nopagebreak
\caption{\label{fig:vnn} The $v^c_d$ and $v^t_d$ in Argonne \argonne\
model are compared with the $v^c_{con}$ and $v^t_{con}$ for the
present $\Psi^N(r_N=0.44$ fm) and for the CKP $\Psi^N$ with
$r_N=0.33$ fm.}
\end{center}
\end{figure}

This study also highlights the need for an improved CQM
Hamiltonian. A primary issue is the inclusion of
relativistic effects. The relativistic kinetic energy,
$\hat{T}=\sum_i \sqrt{\bvec{p}_i^2 + m_i^2}$ is easy to
include. Boosting the two-body potentials has been
studied by Carlson et.\ al.\cite{carlson:rel} and Forest
et.\ al.\cite{forest:rel1} in the context of
light nuclei, and by Isgur \cite{isgur:rel} in a CQM.
The relativization of the current model may have a significant effect
on the six-quark $NN$ state. One consequence of including relativistic
kinetic energy, as shown in CKP, is that with it nucleons have
a smaller rms quark radius. Using the parametrization
for $\Psi^N$ from CKP, we obtain an rms radius
of 0.33 fm, against the 0.44 fm of the present $\Psi^N$.
This has a significant effect on the convolution
potential, Eq.(\ref{eqn:conv}), as seen in Fig.(\ref{fig:vnn}).
As the rms quark radius of the nucleon goes to smaller values the central
term of the convolution potential gets more repulsive near $r\sim 0$
and, more importantly, the tensor gets stronger. Fig.(\ref{fig:vnn})
shows the convolution potential tending toward the Argonne \argonne\
curves as the nucleon quark radius decreases.

The meson exchange picture, correct when nucleons are at large
separation, is questionable when the nucleons overlap considerably.
While the $v^{\pi,LR}_{\sigma\tau}$ and $v^\pi_{t\tau}$
are fixed by the data, the validity of $v^{\pi,SR}_{\sigma\tau}$ is
questionable. As seen from the tables, this term gives a large
repulsive contribution to $\delta E(R_c)$ and one of the ways to get
better agreement with $\delta E_{emp}(R_c)$ may be to reduce its
strength.

\begin{figure}[hb]
\begin{center}
\includegraphics[ width= 340 pt, keepaspectratio, clip ]{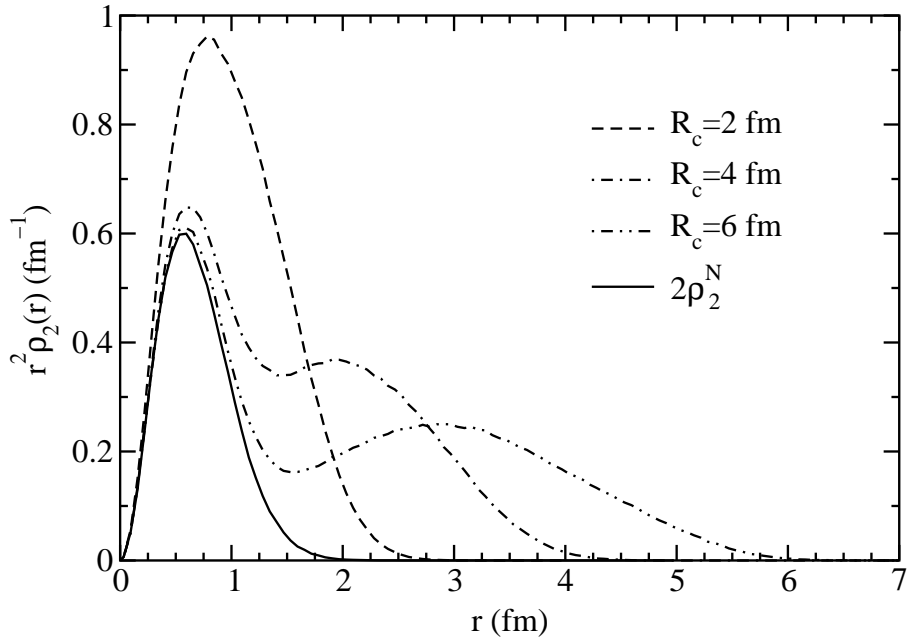}
\nopagebreak
\caption{\label{fig:pdf} Quark pair distribution functions
in six-quark and single nucleon states. The $\rho_2^N$ of the
nucleon is multiplied by a factor of two for easier comparision.}
\end{center}
\end{figure}

Accepting the limitations of the present model and calculations
we can add terms to the six-quark Hamiltonian to reproduce the
$\delta E_{emp}(R_c)$. For example, if we assume quark pairs
interact via a medium range attractive scalar potential $v^S$
of the two-pion exchange range:
\begin{eqnarray}
\label{eqn:tpe}
v^S&=&c^S \sum_{q<q'\le 6} \tilde{T}_\mu^2(r_{qq'}) \\
\tilde{T}_\mu&=&T_\mu - \frac{\Lambda_S^3}{\mu^3} T_{\Lambda_S}
- \frac{1}{2}\frac{\Lambda_S}{\mu}\left(\frac{\Lambda_S^2}{\mu^2}-1\right)
\left(\Lambda_S r+1\right) Y_{\Lambda_S}, \nonumber
\end{eqnarray}
its strength $c^S$ and cutoff $\Lambda_S$ can be adjusted to fit
$\delta E_{emp}(R_c)$. As before $\mu$ is the pion mass.
The two-pion exchange interaction between quarks has recently
been discussed by Riska and Brown \cite{brown99}.
The effect of this potential is calculated perturbatively from
the two-quark distribution functions,
\beq
\label{eqn:pdf}
\rho_2(r) = \frac{\int\: d\bvec{R} \Psi_V^\dagger(\bvec{R})
\sum_{q<q'} \delta^{(3)}\!\left(\rvec - (\rvec_i-\rvec_j)\right)
\Psi_V(\bvec{R})}
{\int d\bvec{R} \Psi_V^\dagger(\bvec{R})\Psi_V(\bvec{R})}
\eeq
which give the probability to find two quarks a distance $r$
apart. They are plotted in Fig.(\ref{fig:pdf}) for
$R_c=2$, 4, and 6 fm. The expectation values of $v^S$ are given by:
\beq
\label{eqn:vs}
\bb v^S \kb = \int d^3r \rho_2(r) v^S(r).
\eeq
The results of this exploratory calculation are shown in
Table \ref{tab:tpe} for $\Lambda_S=5$ fm$^{-1}$ and $c^S$
chosen to reproduce $\delta E_{emp}(R_c=2$ fm).
The $\bb v^S \kb$ in the single nucleon state is $-51.4$ MeV
and $-21.0$ MeV, for $\Lambda=5$ and 2 fm$^{-1}$ cases, respectively,
which is small enough to be treated perturbatively, and it merely
redefines the constant $V_0$. Though the effect of
$v^S$ is minimal on the single hadron spectrum, it is
larger in the six-quark case since there are
nine pairs of quarks in different nucleons. This illustrates
how we could refine the CQM Hamiltonian using $NN$ scattering
data.

The quark pair distribution function in a
single nucleon, $\rho^N_2(r)$, is also plotted in
Fig.(\ref{fig:pdf}), scaled by a factor of 2 for easier comparison
with the $\rho_2$ for six quarks. When $R_c=4$ and
6 fm the first peak in $\rho_2$ at $r\sim 0.6$ fm corresponds to the
distribution of pairs in a single nucleon while the second peak
at larger radii corresponds to the distribution of pairs of
quarks in different nucleons. In these cases the first peak
approximately equals $2\rho^N_2(r)$ as expected. For $R_c=2$ fm
these peaks merge.

\begin{figure}[hb]
\begin{center}
\includegraphics[ width= 375 pt, keepaspectratio, clip ]{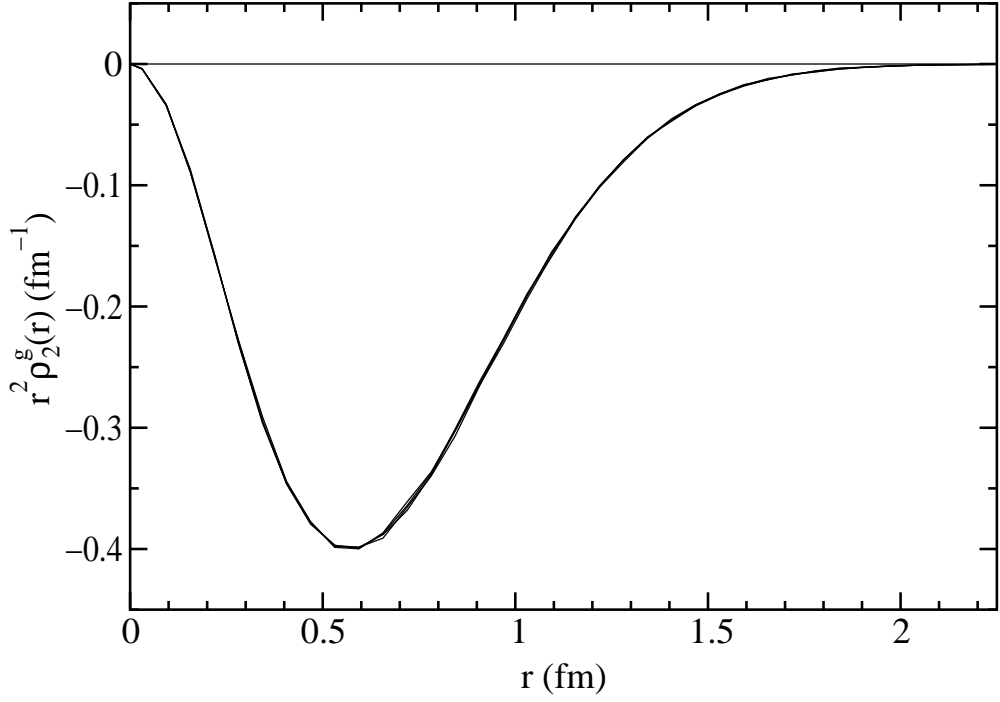}
\nopagebreak
\caption{\label{fig:gpdf} The quark pair
$\mathsf{T}_{q'}\cdot\mathsf{T}_q$ distribution function in
$R_c=2$, 4, and 6 fm six-quark states and in the nucleon. The
nucleon $\rho_2^g(r)$ is multiplied by two for easier comparison.
This figure has four essentially overlapping curves.}
\end{center}
\end{figure}

Further insight into the quark substructure of the two-nucleon states
is gained by calculating the quark-pair distribution
function for the $\mathsf{T}_q\cdot\mathsf{T}_{q'}$ operator
in OGE interactions. It is denoted by, $\rho^g_2$, and defined as:
\beq
\label{eqn:ttpdf}
\rho^g_2(r) = \frac{\int\: d\bvec{R} \Psi_V^\dagger(\bvec{R})
\sum_{q<q'} \mathsf{T}_q\cdot\mathsf{T}_{q'}
\delta^{(3)}\!\left(\rvec - (\rvec_q-\rvec_{q'})\right)
\Psi_V(\bvec{R})}
{\int d\bvec{R} \Psi_V^\dagger(\bvec{R})\Psi_V(\bvec{R})}
\eeq
and plotted in Fig.(\ref{fig:gpdf}). The figure has four
overlapping curves showing $\rho^g_2$ for $R_c=2$, 4, 6 fm and
$2\rho^{g(N)}_2$ in the single nucleon. The
$\mathsf{T}_q\cdot\mathsf{T}_{q'}$ has the value $-2/3$ for quark
pairs single nucleons and is zero for CD terms when the quarks are in
different nucleons. Considering only the CD terms we expect the $\rho^g_2$
to lie on top of each other for all $R_c$. CND contributions can
cause small ($\sim 1\%$) differences in $\rho^g_2$ dependent on $R_c$.
However the accuracy of the present calculation of $\rho^g_2(r)$
is also $\sim 1\%$. Note the absence of a second peak for $r>2$ fm
indicating the lack of long-ranged color-dependent correlations,
as expected.

Although we will not include the results
here, the effective $NN$ wave functions
may be calculated from the $\Psi_V$ using the methods
of Schiavilla et.\ al.\cite{schiavilla} and Forest
et.\ al.\cite{donuts}, and the knowledge of $E(R_c)$
allows one to determine $NN$ scattering phase shifts
\cite{carlson:res}. The aim of the present work was to
explore if QMC calculations can be used to study
the six-quark system with sufficient accuracy to discern
nuclear effects. The present results are obviously
encouraging.

\acknowledgements{The authors would like to thank Steven C. Pieper for
assistance with MPI. This work was supported by National Science Foundation
grant NSF98-00978 and in part by NSF cooperative agreement ACI-9619020
through computing resources provided by the National Partnership for
Advanced Computational Infrastructure at the San Diego Supercomputer
Center.}

\bibliography{6qd}

\pagebreak

\begin{table}[hb]
\begin{center}
\begin{tabular}{c|rrrr}
\hline
$R_c$	&\multicolumn{1}{c}{2 fm} &\multicolumn{1}{c}{4 fm}
        &\multicolumn{1}{c}{6 fm} &\multicolumn{1}{c}{$\infty$} \\
\hline
$E(R_c)/2$	&37.01		&2.32		&$-0.35$	&$-$1.12  \\
$\bb T_N(R_c)\kb/2$
		&79.61		&23.96		&15.16		&9.94    \\
$\bb v_{NN}(R_c)\kb/2$
		&$-$42.62	&$-$21.64	&$-$15.51	&$-11.06$ \\
$E_{NI}(R_c)/2$	&51.16		&12.79		&5.69		&0        \\
$\delta E_{emp}(R_c)/2$
		&$-$14.16	&$-$10.48	&$-$6.04	&$-1.12$  \\ 
\hline
\end{tabular}
\caption{\label{tab:av18} Energies per nucleon calculated from
the isoscalar part of Argonne \argonne\ potential (in MeV) vs.
the diameter of the confining cavity. $R_c\to\infty$ corresponds
to the free deuteron.}
\end{center}
\end{table}

\begin{table}[ht]
\begin{center}
\begin{tabular}{c|rrr|rrr|rr}
\hline
 &\multicolumn{3}{|c|}{$\gft=2$ fm$^{-1}$}
 &\multicolumn{3}{|c|}{$\gft\to\infty$}  \\
$R_c$ &\multicolumn{1}{c}{2 fm} &\multicolumn{1}{c}{4 fm}
      &\multicolumn{1}{c|}{6 fm}&\multicolumn{1}{c}{2 fm}
      &\multicolumn{1}{c}{4 fm} &\multicolumn{1}{c|}{6 fm}
      &\multicolumn{1}{c}{\raisebox{2.5ex}{$N$}}
      &\multicolumn{1}{c}{\raisebox{2.5ex}{$\Delta$}} \\
\hline
$E_V$ &998.9(4) &950.0(4) &943.3(4) &990.7(4) &948.1(4) &942.6(5)
            &938.9(4) &1233.2(4) \\
$T$   &1115(4)  & 1070(4) &1062(4)  & 1118(4) &1065(4)  &1057(4)
            &1053(4)  &755(3)    \\
$V$   &257(4)   &  253(4) &255(4)   &  247(4) &257(4)   & 259(4)
            & 260(4)  &852(2)    \\
\hline
$V^C$ &996(2)   &  995(2) &995(2)   &  997(2) &997(2)   &996(2)
      &996(2)   &1157(2)   \\
\hline
$v^g_c$     
      &$-$340.8(4)&$-$340.9(4)&$-$340.8(4)&$-$340.6(4)&$-$340.3(4)&$-$340.6(4)
      &$-$340.8(4)&$-$308.4(4) \\
$v^g_\sigma$
      &$-$101.8(3)&$-$106.5(4)&$-$107.5(4)&$-$107.0(4)&$-$106.7(4)&$-$106.9(4)
      &$-$107.4(4)&56.6(2)   \\
$v^g_t$
      &$-$4.74(1) &$-$4.31(1)  &$-$4.23(1)&$-$4.32(1) &$-$4.25(1) &$-$4.23(1)
      &$-$4.24(1) &$-$7.75(7)  \\
$v^g_{\ell s}$
      &$-$2.20(1) &$-$2.09(1) &$-$2.06(1) &$-$2.24(1) &$-$2.09(1) &$-$2.07(1)
      &$-$2.08(1) &$-$2.28(5)  \\
\hline
$v^{\pi,SR}_{\sigma\tau}$
      &$-$283(1)  &$-$296(1)  &$-$298(1)  &$-$292(1)  &$-$295(1)  &$-$297(1)
      &$-$299(1)  &$-$37.2(2)  \\
$v^{\pi,LR}_{\sigma\tau}$
      &20.91(3) &22.52(4) &23.13(4) &21.51(3) &22.54(4) &23.10(5)
      &23.69(4) &3.96(1)   \\
$v^\pi_{t\tau}$
      &$-$27.7(1) &$-$14.8(2) &$-$9.7(1)  &$-$24.6(1) &$-$14.4(2) &$-$9.49(2)
      &$-$6.48(1) &$-$9.86(9)  \\
\hline
$E(R_c)$
      & 60.0(6) & 11.1(6)   &  4.4(6)   & 51.8(6) &  9.2(6)   &  3.7(7) \\
$\delta E(R_c)$
      & 8.8(6)  & $-$1.7(6) & $-$1.3(6) &  0.7(6) & $-$3.6(6) & $-$2.0(7) \\
$\delta E_{emp}(R_c)$
      &$-$14.21 &$-$10.5    &$-$6.0     &$-$14.21 &$-$10.5    &$-$6.0\\   
\hline
\end{tabular}
\end{center}
\caption{\label{tab:cfnuc} Energies per nucleon (in MeV) of six-quark
states compared with the single hadron values for single $N$ and $\Delta$.
Statistical errors are shown in parenthesis.}
\end{table}

\begin{table}[ht]
\begin{center}
\begin{tabular}{c|ccc}
\hline
$\mathcal{P}$
 &	    ($ijk$)&	    ($lmn$)&	         $P_{ij}$\\
1&		123&		456&		    ---  \\
2&		423&		156&		 $P_{14}$\\
3&		523&		416&		 $P_{15}$\\
4&		623&		451&		 $P_{16}$\\
5&		143&		256&		 $P_{24}$\\
6&		153&		426&		 $P_{25}$\\
7&		163&		452&		 $P_{26}$\\
8&		124&		356&		 $P_{34}$\\
9&		125&		436&		 $P_{35}$\\
10&		126&		453&		 $P_{36}$\\
\hline
\end{tabular}
\caption{\label{tab:partition}
Six-body wave function partitions and quark exchange operators.
The first column is the partition number, the second and third lists the
quarks joined by flux tubes in singlets, and the last column is
the quark exchange operator which gives that partition when applied
to $\ket{\mathcal{P}=1}$.}
\end{center}
\end{table}

\begin{table}[ht]
\begin{center}
\begin{tabular}{cc}
\hline
\st			&880 MeV \\
\alps			&.61 	 \\
$f_{\pi NN}/4\pi$	&.075	 \\
$m_Q$			&313 MeV \\
$\Lambda$		&5 fm$^{-1}$\\
$\gft$			&2,$\infty$ fm$^{-1}$\\
$V_0$			&373.7 MeV\\
\hline
\end{tabular}
\caption{\label{tab:hpars} CQM Hamiltonian parameters}
\end{center}
\end{table}

\begin{table}[ht]
\begin{center}
\begin{tabular}{c|c|cccc}
\hline
$R_c$ &\multicolumn{1}{|c|}{\gft(fm$^{-1}$)} & $\alpha_c$ & $\alpha_t$
				  & $d_c$ (fm) & $d_t$ (fm)\\
\hline
                         & 2      &.10 &.90 &3.55 &1.25 \\
\raisebox{2.5ex}[0pt]{2} &$\infty$&.17 &.90 &3.35 &1.10 \\
\hline
                         & 2      &.30 &1.0 &10.0 &1.20 \\
\raisebox{2.5ex}[0pt]{4} &$\infty$&.50 &1.0 &10.0 &1.10 \\
\hline
                         & 2      &.90 &.93 &10.0 &1.20 \\
\raisebox{2.5ex}[0pt]{6} &$\infty$&.34 &.72 &10.0 &1.20 \\
\hline
\end{tabular}
\caption{\label{tab:params} Optimum values of the variational
parameters.}
\end{center}
\end{table}

\begin{table}[ht]
\begin{center}
\begin{tabular}{c|rrr|r|rrr}
\hline
 & \multicolumn{3}{|c|}{$\gft=2$ fm$^{-1}$} & $\gft\to\infty$
 & \multicolumn{3}{|c}{$\gft=0$} \\
            &\multicolumn{1}{c}{CD}&\multicolumn{1}{c}{CND}
	    &\multicolumn{1}{c|}{Total}&\multicolumn{1}{c|}{Total}
            &\multicolumn{1}{c}{CD}&\multicolumn{1}{c}{CND}
	    &\multicolumn{1}{c}{Total} \\
\hline
$E_V$       &990.8(8)   &8.1(3)    &998.9(4)    &990.7(4)    &990.3(8)   &25.1(3) &1015.0(6) \\
$T$	    &1117(4)    &$-$2(1)   &1115(4)     &1117(4)     &1113(4)    & 7(1)   &1119(4)   \\
$V$	    &247(4)     &10.4(2)   &257(4)      &247(4)      &252(3)     &18.1(3) &270(3)    \\
\hline
$V^C$       &997(2)     &$-$0.44(3)&996(2)      &997(2)      &1000(2)    &2.1(4)  &1002(2)   \\
\hline
$v^g_c$     &$-$340.5(4)&$-$.2(2)  &$-$340.8(4) &$-$340.6(4) &$-$340.0(4)&$-$.5(3) &$-$340.4(4) \\
$v^g_\sigma$&$-$106.2(3)&4.4(2)    &$-$101.8(3) &$-$107.0(4) &$-$106.0(3)&8.0(3)   &$-$98.0(3)  \\
$v^g_t$     &$-$4.31(1) &$-$.43(1) &$-$4.74(1)  &$-$4.33(1)  &$-$4.30(1) &$-$1.40(3)&$-$5.69(2)  \\
$v^g_{\ell s}$
	    &$-$2.26(1) &.06(1)    &$-$2.20(1)  &$-$2.24(1)  &$-$2.25(1) &.024(6)  &$-$2.22(1)  \\
\hline
$v^{\pi,SR}_{\sigma\tau}$
	    &$-$290(1)  & 8.1(5)   &$-$283(1)   &$-$292(1)   &$-$290(1)  &12.9(4)    &$-$277(1)   \\
$v^{\pi,LR}_{\sigma\tau}$
	    &21.47(3)   &$-$.56(3) &20.91(3)    &21.52(3)    &21.44(4) &$-$1.71(6) &19.71(4)  \\
$v^\pi_{t\tau}$
	    &$-$27.35(1)&$-$.39(1) &$-$27.7(1)  &$-$24.6(1)  &$-$27.3(1) &$-$1.29(3)&$-$28.6(1)  \\
\hline
\end{tabular}
\end{center}
\caption{\label{tab:rc2}
Variationally determined six-quark state energies for cavity
diameter $R_c=2$ fm for finite ($\gft=2$ fm$^{-1}$) and strong
($\gft\to\infty$) coupling. The columns give color-diagonal,
CD, color-non diagonal, CND and total contributions. The $\gft=0$
state is evaluated perturbatively.}
\end{table}

\begin{table}[ht]
\begin{center}
\begin{tabular}{c|rrr|r}
\hline
 & \multicolumn{3}{|c|}{$\gft=2$ fm$^{-1}$} & $\gft\to\infty$ \\
            &\multicolumn{1}{c}{CD}&\multicolumn{1}{c}{CND}
	    &\multicolumn{1}{c|}{Total}&\multicolumn{1}{c}{Total} \\
\hline
$E_V$       &948.8(5)   &1.2(1)     &950.0(4)     &948.1(4)  \\
$T$	    &1070(4)    &$-$.3(4)   &1070(4)      &1065(4)   \\
$V$	    &252(4)     &1.5(1)     &253(4)       &257(4)    \\
\hline
$V^C$       &995(2)     &.0(1)      &995(2)       &997(2)    \\
\hline
$v^g_c$     &$-$340.8(4)&$-$.04(9)  &$-$340.9(4)  &$-$340.3(4) \\
$v^g_\sigma$&$-$107.1(4)&.6(1)      &$-$106.5(4)  &$-$106.7(4) \\
$v^g_t$     &$-$4.25(1) &$-$.06(1)  &$-$4.31(1)   &$-$4.25(1)  \\
$v^g_{\ell s}$
	    &$-$2.10(1) &.002(1)    &$-$2.09(1)   &$-$2.09(1)  \\
\hline
$v^{\pi,SR}_{\sigma\tau}$
	    &$-$296(1)  &1.1(2)     &$-$296(1)    &$-$295(1)   \\
$v^{\pi,LR}_{\sigma\tau}$
	    &22.59(4)   &$-$.07(1)  &22.52(3)     &22.54(3)  \\
$v^\pi_{t\tau}$
	    &$-$14.8(2) &$-$.04(1)  &$-$14.8(1)   &$-$14.4(2)  \\
\hline
\end{tabular}
\caption{\label{tab:rc4}
Six-quark state energies for cavity diameter $R_c=4$ fm.
Notation identical to Table \ref{tab:rc2}.}
\end{center}
\end{table}

\begin{table}[ht]
\begin{center}
\begin{tabular}{c|rrr|r}
\hline
 & \multicolumn{3}{|c|}{$\gft=2$ fm$^{-1}$} & $\gft\to\infty$ \\
            &\multicolumn{1}{c}{CD}&\multicolumn{1}{c}{CND}
	    &\multicolumn{1}{c|}{Total}&\multicolumn{1}{c}{Total} \\
\hline
$E_V$       &943.1(5)   &.27(3)     &943.3(2)   &942.6(5)  \\
$T$	    &1062(4)    &.0(1)      &1062(4)    &1057(4)   \\
$V$	    &255(4)     &.23(3)     &255(4)     &259(4)    \\
\hline
$V^C$       &995(2)     &$-$.01(5)  &995(2)     &996(2)    \\
\hline
$v^g_c$     &$-$340.8(4)&.02(3)     &$-$340.8(4)&$-$340.6(4) \\
$v^g_\sigma$&$-$107.6(4)&.12(4)     &$-$107.5(4)&$-$106.9(4) \\
$v^g_t$     &$-$4.22(1) &$-$.043(1) &$-$4.23(1) &$-$4.23(1)  \\
$v^g_{\ell s}$
	    &$-$2.06(1) &.002(1)    &$-$2.06(1) &$-$2.07(1)  \\
\hline
$v^{\pi,SR}_{\sigma\tau}$
	    &$-$300(1)  &.18(8)     &$-$298(1)  &$-$297(1)   \\
$v^{\pi,LR}_{\sigma\tau}$
	    &23.14(4)   &$-$.01(1)  &23.13(4)   &23.10(3)  \\
$v^\pi_{t\tau}$
	    &$-$9.6(1)  &$-$.007(2) &$-$9.7(1)  &$-$9.49(2)  \\
\hline
\end{tabular}
\caption{\label{tab:rc6}
Six-quark state energies for cavity diameter $R_c=6$ fm.
Notation identical to Table \ref{tab:rc2}.}
\end{center}
\end{table}

\begin{table}[ht]
\begin{center}
\begin{tabular}{cc|rrrr}
\hline
$c^S (MeV)$ & $\Lambda_s$ (fm$^{-1}$)
&\multicolumn{1}{c}{2 fm}
&\multicolumn{1}{c}{4 fm}
&\multicolumn{1}{c}{6 fm} \\
\hline
0	 &      &8.8(4)    &$-$1.7(4) &$-$1.3(4) \\
$-$0.0077&  5	&$-$14.2(4)&$-$6.5(4) &$-$2.6(4) \\
$-$0.67	 &  2	&$-$14.2(4)&$-$11.0(4)&$-$5.1(4) \\
\multicolumn{2}{c|}{$\delta E_{emp}$}
		& $-$14.2& $-$10.5    &$-$6.0    \\
\hline
\end{tabular}
\caption{\label{tab:tpe} The $\delta E(R_c)$ for various $c^S$ and
$\Lambda_S$ compared with $\delta E_{emp}(R_c)$ from Argonne
\argonne\ potential.}
\end{center}
\end{table}

\clearpage

\end{document}